\documentclass[aps,prd,twocolumn,showpacs]{revtex4-1}
\pdfoutput=1 
\usepackage{atlasphysics}
\usepackage{graphicx}
\usepackage{dcolumn}
\usepackage{bm}
\usepackage{url}

\usepackage{multirow,bigdelim}
\usepackage{subfigure}
\usepackage{atlasphysics}
\usepackage{booktabs} 
\usepackage{epstopdf}
\usepackage[section]{placeins}
\usepackage[hyperindex,breaklinks]{hyperref}

\newcommand{\etall}{\ensuremath{\eta^{\ell\ell}}}
\newcommand{\ptll}{p_{\rm T}^{\ell\ell}}

\newcommand{\dphimetz}{\ensuremath{\Delta\phi(\vec{E}^{\rm miss}_{\rm T}, \ptll)}}

\newcolumntype{x}[1]{>{\centering\hspace{0pt}}p{#1}}
\newcolumntype{a}[1]{>{\raggedright\hspace{0pt}}p{#1}}
\newcommand{\tn}{\tabularnewline}

\def\thetitle{Search for dark matter in events with a $Z$ boson and missing transverse momentum in $pp$ collisions at $\sqrt{s}=8$~TeV with the ATLAS detector}

\usepackage{preprintcover}  
\PreprintCoverPaperTitle{Search for dark matter in events with a $Z$ boson and missing transverse momentum in $pp$ collisions at $\sqrt{s}=8$~TeV with the ATLAS detector}  
\PreprintIdNumber{CERN-PH-EP-2013-231}  
\PreprintCoverAbstract{
A search is presented for production of dark matter particles recoiling against 
 a leptonically decaying $Z$ boson in 20.3 fb$^{-1}$ of $pp$ collisions at $\sqrt{s}=8$~TeV 
 with the ATLAS detector at the Large Hadron
 Collider. Events with large missing transverse momentum and two 
 oppositely-charged electrons or muons consistent with
 the decay of a $Z$ boson are analyzed. No excess above the 
 Standard Model prediction is observed. 
 Limits are set on the mass scale of the contact interaction as a function of the  
 dark matter particle mass using an effective field theory 
 description of the interaction of dark 
 matter with quarks or with $Z$ bosons. Limits are also set on the coupling and mediator mass of a 
 model in which the interaction is mediated by a scalar particle. 
}  
\PreprintJournalName{PRD} 

\begin{document}

\title{\thetitle}

\date{\today}

\begin{abstract}
A search is presented for production of dark matter particles recoiling against 
 a leptonically decaying $Z$ boson in 20.3 fb$^{-1}$ of $pp$ collisions at $\sqrt{s}=8$~TeV 
 with the ATLAS detector at the Large Hadron
 Collider. Events with large missing transverse momentum and two 
 oppositely-charged electrons or muons consistent with
 the decay of a $Z$ boson are analyzed. No excess above the 
 Standard Model prediction is observed. 
 Limits are set on the mass scale of the contact interaction as a function of the  
 dark matter particle mass using an effective field theory 
 description of the interaction of dark 
 matter with quarks or with $Z$ bosons. Limits are also set on the coupling and mediator mass of a 
 model in which the interaction is mediated by a scalar particle. 
\end{abstract}


\maketitle

\clearpage

Astrophysical measurements indicate the existence of
non-baryonic dark matter~\cite{rotationcurve,wmap}.  
However, collider based searches, nuclear scattering experiments, and searches 
for particles produced from dark-matter annihilation have not yet
revealed its particle nature nor discovered its non-gravitational
interactions, if they exist~\cite{pdg}. 
Collider-based searches for weakly interacting massive particles (WIMPs, denoted as $\chi$),  
specifically $pp\rightarrow\chi\bar{\chi}+X$ at the Large Hadron Collider (LHC) via some unknown
intermediate state, are an important facet of the experimental
program and provide sensitivity over a broad range of values of the WIMP mass, $m_{\chi}$, 
including for low masses  
where direct detection experiments are less sensitive.
The presence of dark-matter particles, not directly observable in a collider detector, can 
be inferred from their recoil against Standard Model (SM) particles. 
The LHC collaborations have reported limits on the cross section 
for the process that includes initial state radiation (ISR),  
$pp\rightarrow\chi\bar{\chi}+X$, where the ISR component $X$ is a 
hadronic jet~\cite{atlasjet,cmsjet}, a 
photon~\cite{atlasphoton,cmsphoton}, or a $W$ or $Z$ boson decaying 
hadronically~\cite{atlaswz}. Limits on 
dark matter produced in the decay of the Higgs boson 
have also been reported~\cite{ZHinv}. In this analysis, limits are set using 
the final state of a $Z$ boson decaying to two oppositely charged electrons or 
muons, plus missing transverse momentum, \met.

Since the nature of the intermediate state mediating the parton--WIMP interaction is not known, a useful
approach is to construct an effective field
theory (EFT)~\cite{Goodman:2010ku,Beltran:2010ww,Fox:2011pm}.
EFTs have often been used to describe
interactions between dark-matter particles and quarks or gluons,
but they have recently been extended to describe direct interactions
with electroweak bosons~\cite{monoz7tev,Cotta:2012nj,axionZ}.
In the context of the EFT framework, the WIMP is considered to be the only new particle accessible at LHC energies, in addition to
the SM fields. The 
mediator of the interaction is assumed to be heavy compared to the typical parton interaction energies
involved, and 
the dark-matter particles are also assumed to be produced in pairs. 

\begin{figure}
\begin{center}
\subfigure[]{\includegraphics[width=36mm]{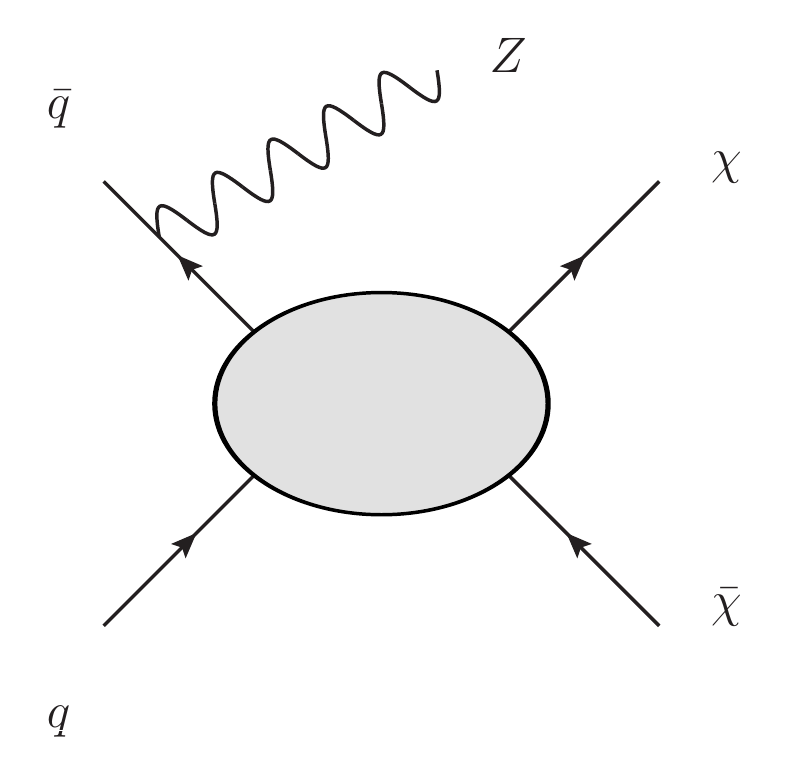}}
\subfigure[]{\includegraphics[width=48mm]{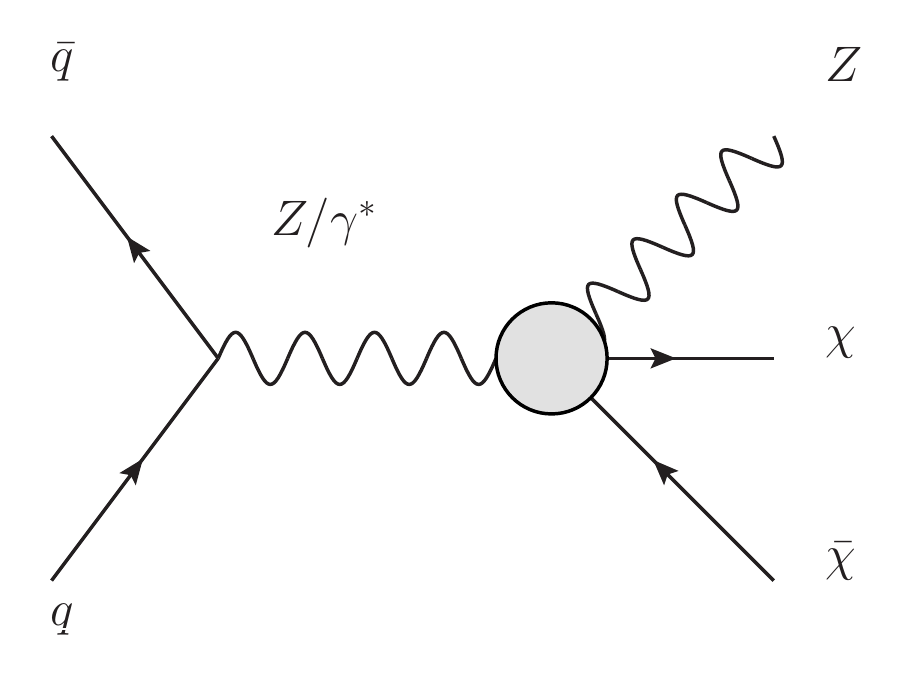}}
\caption{\label{fig:diag_monoz} The diagrams showing different types of $pp\rightarrow\chi\bar{\chi}+Z$ production modes  
considered in this analysis~\cite{monoz7tev}. Figure (a) shows a diagram that includes an ISR operator, and figure (b) shows a diagram that includes a $ZZ\chi\chi$ operator.}
\end{center}
\end{figure}

The EFTs considered in this analysis, depicted in Fig.~\ref{fig:diag_monoz}, are expressed in terms of two parameters:
$m_{\chi}$ and a mass scale, $M_{\star}$, described in Ref.~\cite{Goodman:2010ku}.
$M_{\star}$ parameterizes the coupling between the WIMP and SM particles,  
where the coupling strength is normalized, or in inverse proportion, to the heavy-mediator mass scale.
The coefficients of the Lagrangian's interaction terms
appear as powers of $M_{\star}$, e.g. for the D1 operator as $1/M_{\star}^3$ and 
for the D5 and D9 operators as $1/M_{\star}^2$. 
The definition of the D1, D5, and D9 operators and the region of validity of the EFT limits are discussed in Ref.~\cite{Goodman:2010ku, eftValid}.

Following the approach of Ref.~\cite{monoz7tev}, the coupling
of dark matter to electroweak bosons is considered for dimension-5 and dimension-7 operators.
The dimension-7 operator couples dark matter to $Z\gamma^{*}$ as
well as $ZZ$. Since a $Z$ boson is in the final state for each
operator,  intermediate states with a $Z$ or $\gamma^{*}$ each contribute
to the matrix element. The relative contribution of the $Z$ and
$\gamma^{*}$ diagrams is a parameter of the theory.

This analysis considers models of dark-matter production where a $Z$ boson is radiated as ISR
or interacts directly with WIMPs. 
The latter case of an interaction between a $Z$-boson and a WIMP is 
a process not previously investigated in the analysis of LHC experiments.

To complement the EFT analysis, this paper also examines the results in terms of a model in which the
intermediate state is specified~\cite{bell}. 
In this model a scalar-mediator $\eta$, with mass $m_{\eta}$, and a scalar--WIMP coupling strength $f$
is responsible for the production of the dark-matter particles. 
The mediator $\eta$ transforms as a color triplet and an electroweak doublet, and 
has a hypercharge of $1/3$. 
The production cross section is proportional to $f^4$. 
The same final state signature predicted by the EFT, $Z\chi\bar{\chi}$, is produced. This 
process is similar to SUSY processes in which the 
$\chi$ is a neutralino and the $\eta$ is a squark doublet, but without a 
direct gluino analogue. This corresponds to a scenario where the gluino would be too heavy to be produced at the LHC 
and is therefore irrelevant. 

In this analysis, a search for the production of a $Z$ boson 
with subsequent decay to $e^+e^-$ or $\mu^+\mu^-$ in 
association with large \MET\ from the escaping 
$\chi\bar{\chi}$ particles is reported, based on 20.3~fb$^{-1}$ 
of $pp$ collision data collected by the ATLAS detector at a 
center-of-mass energy of $\sqrt{s} = 8$~TeV. 
Only data collected with stable beams and all detector subsystems fully operational are used. 
Several signal regions with different requirements on the 
\MET\ are defined to best probe the variety of 
models tested.

The ATLAS detector~\cite{atlas_det} consists of an
inner detector (ID) surrounded by a solenoid that produces a 2~T magnetic field,
electromagnetic and hadronic calorimeters, and a muon spectrometer (MS) 
employing toroidal magnets. The ID measures charged-particle tracks 
over the full azimuthal angle and in a 
pseudorapidity~\footnote{
ATLAS uses a right-handed coordinate system with its origin at the nominal 
interaction point (IP) in the center of the detector and the $z$-axis 
along the beam pipe. The $x$-axis points from the IP to the center of the 
LHC ring, and the $y$-axis points upward. Cylindrical coordinates $(r,\phi)$ 
are used in the transverse plane, $\phi$ being the azimuthal angle around 
the beam pipe. The pseudorapidity is defined in terms of the polar angle 
$\theta$ as $\eta=-\ln\tan(\theta/2)$.
}
range of $|\eta|<2.5$  
using silicon pixel, silicon microstrip, and transition-radiation straw-tube detectors, 
the last of which also distinguishes electrons from heavier charged particles in the range $|\eta| < 2.0$.
Liquid-argon (LAr) electromagnetic sampling calorimeters cover the range $|\eta| <
3.2$ with a typical granularity in $\Delta\eta \times \Delta\phi$ of
$0.025 \times 0.025$.  A scintillator-tile calorimeter
provides hadronic calorimetry for $|\eta| < 1.7$. In the endcaps ($|\eta| > 1.5$), LAr
is also used for the hadronic calorimeters matching the 
outer $|\eta|$ limit of the endcap electromagnetic calorimeters. The LAr 
forward calorimeters extend the coverage to $|\eta|<4.9$ and provide both the electromagnetic and hadronic energy 
measurements. 
The MS covers $|\eta|<2.7$ and provides triggering and 
precision tracking for muons. 
A three-level trigger system 
is used to select interesting events to be recorded for subsequent 
offline analysis.

Electrons are required to have 
transverse energy, $E_{\rm T}$, larger than $20 \GeV$ and $|\eta|<2.47$. 
The $E_{\rm T}$ is measured from the energy deposited in the 
electromagnetic calorimeter, and the electron's direction from the ID track. 
Electrons must satisfy the {\it medium} object quality requirements from Ref.~\cite{egammanote} 
updated for 2012 run conditions,
which are based on calorimeter 
shower shape, ID track quality, and the spatial match between the shower and 
the track. 
Electrons must be isolated, 
satisfying 
$\sum^{\Delta{R}<0.2} \pT^{\rm track}/E_{\rm T}<0.1$, where the sum is over the 
transverse momenta, $\pT^{\rm track}$, of all other ID tracks with $\pT^{\rm track}>1 \GeV$ within a cone 
of radius $\Delta{R}=\sqrt{\left(\Delta\eta\right)^{2} + \left(\Delta\phi\right)^{2}}=0.2$ around the electron direction. 

Muons are required 
to have $\pT>20$ GeV and $|\eta|<2.5$. 
A combined fit of the ID and MS tracks is used to 
reconstruct the muon \pT. 
High-quality tracks are ensured by requirements on the number of 
hits in the ID. 
Longitudinal and transverse impact parameters, $z_{0}$ and $d_{0}$ respectively, must satisfy 
 $|z_{0}|<10$ mm and $|d_{0}|<1$ mm, with respect to the primary vertex, 
defined as the vertex with the highest $\sum \left(\pt^{\rm track}\right)^2$. 
The muon must be isolated, satisfying  $\sum^{\Delta{R}<0.2} \pT^{\rm track}/\pT<0.1$; 
here again, the muon track itself is excluded from the sum.

The anti-$k_{t}$ jet algorithm~\cite{ktjet} with radius parameter of 0.4 is used to reconstruct jets from 
topological clusters~\cite{cluster}, which are three-dimensional clusters of neighboring energy 
deposits in the calorimeter cells.
A calibration procedure is used in which 
the raw energy  measurement from the calorimeter cluster is corrected to the jet energy scale. 
Jets are required to have $\pT>25 \GeV$ and $|\eta|<2.5$. 
Jets from secondary proton--proton collisions are removed by requiring that most of 
the tracks associated with the jet, weighted by \pt,  
originate at the primary vertex. 

Since muons may generate delta-ray electrons or radiate photons that produce electron--positron pairs, 
electrons closer than 
$\Delta R=0.2$ to a muon that passes the
analysis selection are rejected. In addition, electrons and muons closer than 
$\Delta R=0.4$ to a jet 
are also rejected. 

The measurement of the missing transverse momentum, a 
vector in the transverse plane, \vecMET, with magnitude \MET, is 
based on the measurement of the energy collected by the calorimeters and the 
momenta of muons. 
Muons and 
electrons with $\pT>10~\GeV$, jets with $\pT>20~\GeV$, low-\pT\ tracks which don't seed a 
topological cluster, and 
topological clusters not associated with a jet are included in the \vecMET\ calculation~\cite{metnote}.

The candidate signal events were accepted by at least one of the several triggers 
that require either two leptons with low \pT\ or a single lepton 
with higher \pT.   An event must have at least one reconstructed 
vertex with at least three associated tracks with $\pT>400 \MeV$ 
to remove non-collision background. In addition, events 
must have two oppositely charged electrons or muons 
with invariant mass $m_{\ell\ell} \in [76,106]$ GeV to form a $Z$ boson candidate. 
In order to suppress events where the \MET\ originates from mismeasured
jets, the azimuthal angle  between the dilepton system and the \vecMET, $\dphimetz$, must be greater than
2.5, the absolute value of the pseudorapidity of the dilepton system, $|\etall|$, must be less than 2.5, 
and the ratio $|\ptll-\MET|/\ptll$ must be less than 0.5, where $\ptll$ is the transverse momentum of 
the dilepton system. 
Events are removed if they contain one or more jets with $\pT>25 \GeV$ to suppress top-quark pair background. Similarly, 
events containing a third lepton with $\pt>7 \GeV$, satisfying looser identification requirements than invoked for 
the leptons produced in the decay of the $Z$ boson,  
are removed to suppress diboson background. 

The various dark-matter models considered here
have different \MET\ spectra, leading to a variety of optimal lower thresholds of \met. 
Four inclusive signal regions are defined 
with lower thresholds in \MET\ of
150, 250, 350, and 450~GeV.  

The dominant background process is 
$ZZ\rightarrow\ell^+\ell^-\bar{\nu}\nu$ ($\ell=e$, $\mu$), an irreducible background. 
The other irreducible background is $WW\rightarrow\ell^+\nu\ell^-\bar{\nu}$, which 
may only be reduced through the mass window requirement. 
Reducible backgrounds may have 
jets produced in association with two leptons, where the jets are misidentified or 
unreconstructed, such as 
$t\bar{t}\rightarrow \ell^+\nu b\ell^-\bar{\nu}\bar{b}$, 
$ZZ\rightarrow\ell^+\ell^- \bar{q}q$, 
$WZ\rightarrow \bar{q}q'\ell^+\ell^- $, or 
$Z$+jets (including decays to $\tau$ leptons).
Additional reducible sources may produce events with more than two leptons 
where the additional leptons are misidentified or not reconstructed, or 
less than two leptons where jets are misidentified as leptons, such as 
$WZ\rightarrow \ell\nu\ell^+\ell^-$ and  
$W$+jets, respectively.

The $WZ$ and $ZZ$ backgrounds are estimated from Monte Carlo (MC) simulation. 
The next-to-leading-order MC generator
POWHEG~BOX~1.0~\cite{powhegbox} is used, which models the production from a
$q\bar{q}$ initial state. 
The $ZZ$ background sample is normalized to include the $gg\rightarrow ZZ$ contribution using
MCFM~6.2~\cite{mcfm}.
Parton distribution functions (PDFs) for these
samples are modeled using CT10~\cite{ct10pdf}. 
The underlying event, as well as parton showers and hadronization, are simulated with 
PYTHIA~8.165~\cite{pythia8}, using the AU2 tune~\cite{pythia_au2}. 
QED radiative corrections are calculated
using PHOTOS++~3.0~\cite{photos}. 
The detector response is simulated with GEANT~4~\cite{simulation,geant4}.
The simulated samples are generated with pileup conditions
similar to those observed in data. Pileup refers to the multiple interactions occurring in the 
same, or adjacent, beam bunch crossings as the hard process. 
Simulated events are reweighted so that the number of pileup interactions has 
the same distribution as in data events. 

The $WW$, $\ttbar$, $Wt$, and $Z\rightarrow\tau\tau$ backgrounds are estimated from 
data using the absence of signal in the $e\mu$ channel and the relative production 
rate of 1:1:2 for the $ee$, $\mu\mu$, and $e\mu$ channels. 
An $e\mu$ control region similar to the signal region is defined, and the background 
estimate for the $ee$ and $\mu\mu$ signal regions is obtained from the number 
of $e\mu$ events in the control region after correcting 
for the slightly different electron and muon acceptances and efficiencies. 

The $Z$+jets background is estimated using two data-driven techniques. 
The first method, commonly referred to as the ABCD method~\cite{ABCD}, 
considers the distribution of signal and background events in a 
phase space defined by two uncorrelated variables, here \MET\ and \etall, 
for which the signal and background 
have different shapes. 
The phase space is partitioned into four regions labeled A, B, C, and D. Region A is the signal region 
where selection requirements on both variables are invoked, while regions B, C, and D are control regions in which one or both selections are 
reversed. Contamination by signal events in the control regions is found to be negligible. 
The number of events in one control region scaled by the ratio of background events in two other 
control regions estimates the background contribution in the signal region. 
In the second method, the contribution is measured by fitting the distributions of
$\dphimetz$ and \MET\ at small values ($\MET<80$~GeV and $\dphimetz<2.5$) 
and extrapolating them to the signal regions. The two methods invoke all the standard 
selection requirements and give consistent results. 
The ABCD method is used to provide the background estimate, and the difference between 
the two is taken as a systematic uncertainty on the estimate.

The $W$+jets background is estimated by reversing the electron isolation condition 
and loosening identification requirements for one electron in order to obtain a data sample enriched in jets reconstructed as electrons. 
The resultant \MET\ distribution is fitted with a function  of the form $N=A \cdot \left(\MET\right)^b$ 
below 300~GeV and extrapolated to the highest 450~GeV signal region.  
The fitted function is integrated over \MET\ to obtain an estimated background above a given \MET\ threshold. 
A normalization factor is derived from data in the low-\MET\ region with all the analysis 
selections applied. This factor is applied to the extrapolated result to obtain an estimate of this background.

Background estimates are  validated in signal-depleted control regions that are determined by similar
selection criteria used to define the signal region, but where a requirement may be inverted or modified.
For the dominant $ZZ\rightarrow\ell^+\ell^-\bar{\nu}\nu$  background, estimated with MC simulation, the control region probes 
four-lepton events and is defined by
the presence of two pairs of same-flavor, oppositely-charged leptons for which the invariant mass of the pairs 
is within the $Z$ boson mass window. For the subdominant $WZ\rightarrow \ell\nu\ell^+\ell^-$ background,
the control region is characterized  by three leptons: a pair of same-flavor, oppositely-charged leptons for which the invariant 
mass is within the $Z$ boson mass window, an additional electron or muon, and a reconstructed \MET $ > 80$ GeV.
For the minor $WW$ and top-quark background, derived from a data-driven technique, events containing an  
electron and a muon ($e\mu$) with opposite charge are used.
The expected signal region event yield is
obtained from correction factors that account for the relative dilepton reconstruction efficiencies and the ratio of 
same-flavor lepton production
to mixed flavor. 
The predicted yields from MC simulation are consistent with the data-driven estimates and in all cases are consistent with the control region yields observed in data.

Samples of $pp\rightarrow Z\chi\bar{\chi}$ signals are generated using
MadGraph~5~1.5.2~\cite{madgraph5} with parton showering and hadronization modeled
by PYTHIA~8.170 using the MSTW2008 leading-order PDFs~\cite{mstw2008} and the AU2
tune. EFT operators D1 (scalar, spin independent), D5 (vector, spin independent), and
D9 (tensor, spin-dependent), following the definitions of
Ref.~\cite{Goodman:2010ku}, are representative of the full
set of operators in which the $Z$ boson is emitted as ISR.
Similar
\met\ distributions result from all the operators within each of the
three types: scalar, vector, and tensor. 

Two examples of the dimension-7 $ZZ\chi\chi$ operator mixtures 
are considered: one in which the $Z\gamma^{*}$ contribution is
negligible and one in which it is maximal.  Therefore the dimension-7 operators are referred to as
$ZZ\chi\chi$-maximal-$\gamma^{*}$ and $ZZ\chi\chi$-no-$\gamma^{*}$ while the dimension-5 operator
is referred to simply as $ZZ\chi\chi$. 
Cross sections for a few representative operators and for the
scalar-mediator theory with representative coupling constant, $f=6$, and
$m_{\eta}= 1$~TeV are given in Table~\ref{tab:xsec_theory}.

\begin{table}[htbp]
\begin{center}
\caption{\label{tab:xsec_theory} The power dependence of $1/M_{\star}$ for the EFT and the 
cross sections of WIMP production in association with an  
on-shell $Z$ boson for various EFT operators and the scalar-mediator theory are shown. 
For the calculation of the production cross section, $M_{\star}$ is taken to be 1 TeV for the EFT operators.  
The coupling constant of the scalar-mediator theory, $f$, is taken to be 6 and the mass of the 
mediator, $m_{\eta}$, is 1 TeV. }
\begin{tabular}{x{2.2cm}x{1.3cm}x{1.3cm}x{1.4cm}x{1.5cm}}
\hline
\hline
\multirow{2}{*}{} & D5 & D9 & $ZZ\chi\chi$ max. $\gamma^{*}$ & Scalar mediator \tn
\hline
$m_{\chi}$ [GeV] & \multicolumn{4}{c}{Cross sections [fb]} \tn
1    & 7.2  & 130 & 3.1   & 830 \tn
10   & 7.1  & 120 & 3.1   & 810 \tn
200  & 5.6  & 89  & 2.0   & 300 \tn
400  & 3.1  & 47  & 0.83  & 70 \tn
1000 & 0.25 & 3.4 & 0.023 & - \tn
\hline
$M_{\star}^{-1}$ power & 2 & 2 & 3 & - \tn
\hline
\hline
\end{tabular}
\end{center}
\end{table}

\begin{figure}
\begin{center}
\includegraphics[width=0.45\textwidth]{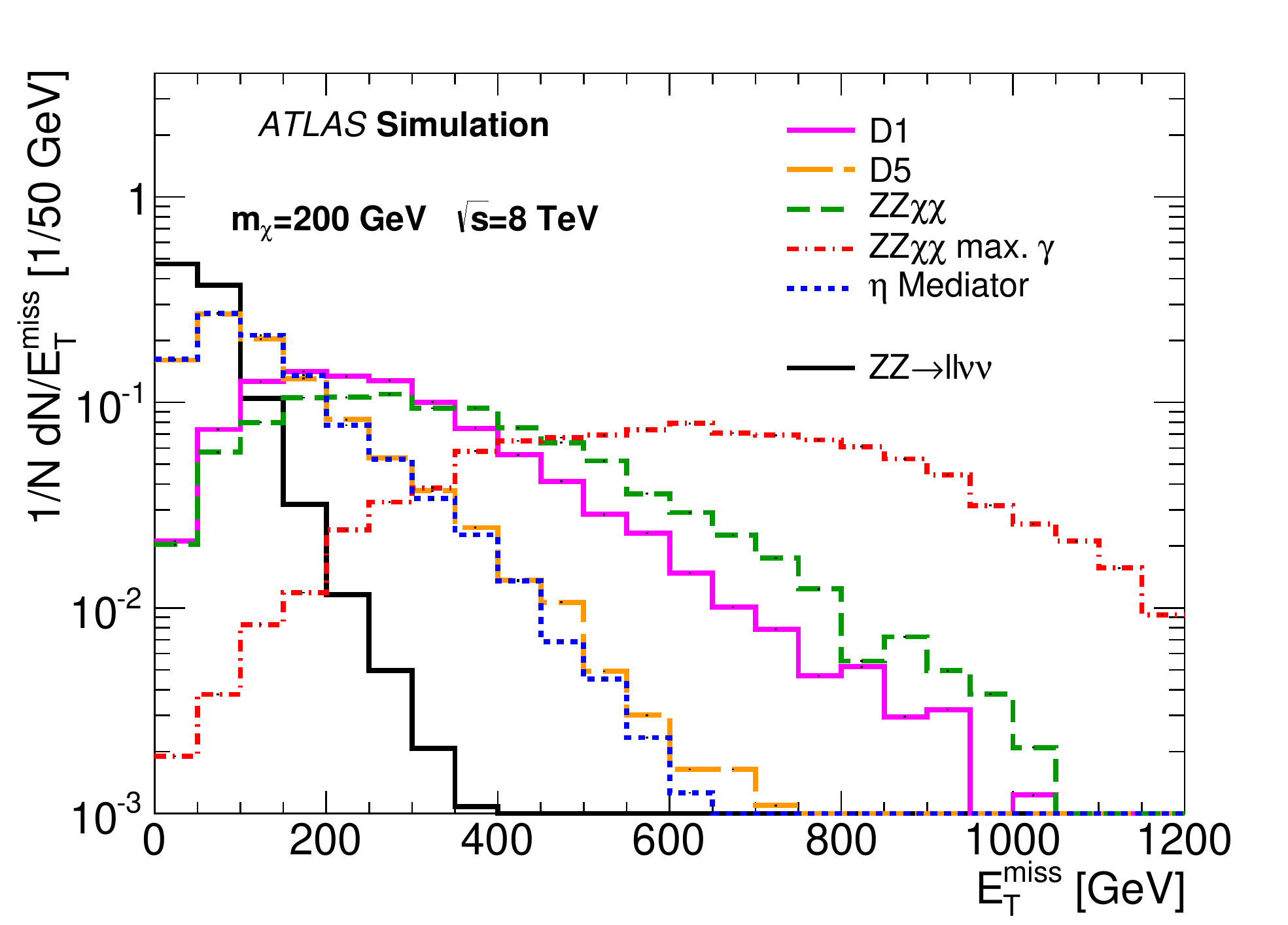}
\caption{\label{fig:met_operator}Shape of the \MET\ distribution in simulated
samples of $ZZ$ background, effective field theories of dark-matter
interaction with a $q\bar{q}$ initial state (D1, D5, and 
D9~\cite{Goodman:2010ku}) and interaction with a $Z/\gamma^{*}$
intermediate state~\cite{monoz7tev}, and the scalar-mediator theory. 
The shapes of $ZZ\chi\chi$-no-$\gamma^{*}$ and $ZZ\chi\chi$-maximal-$\gamma^{*}$ are similar, 
as are the shapes of D9 and the dimension-5 $ZZ\chi\chi$ EFT, so only one of each 
is plotted. 
Each distribution is normalized to unit area. 
The mass of the scalar mediator, $m_{\eta}$ is 1 TeV, 
and the dark-matter particle mass is $m_{\chi}=200\GeV$. }
\end{center}
\end{figure}

Samples of $pp\rightarrow Z\chi\bar{\chi}$ events are 
propagated through the ATLAS detector using the full simulation of the ID and muon trackers and the 
parameterized simulation of the calorimeter~\cite{simulation}, 
tuned to full simulation and data. 
The shapes of the simulated \MET\ distributions 
for the signal operators are shown in Fig.~\ref{fig:met_operator} compared to the dominant 
SM background process $ZZ\rightarrow\ell^+\ell^-\bar{\nu}\nu$. 

Contributions to the systematic uncertainty of the expected SM backgrounds are due largely to experimental sources
affecting the \MET\ measurement and to the efficiencies for the reconstruction and identification of electrons and muons.
For example, when $\MET>$120~GeV, the experimental systematic uncertainty for the $ZZ$ background is 
dominated by the jet--energy scale  
(1.7\% and 2.3\% for electron and muon final states, respectively) and the electron and muon momentum scale 
(2.3\% and 0.8\%, respectively). 
Smaller systematic uncertainties are associated with the \MET\ measurement and with the efficiencies for the reconstruction 
and identification of electrons and muons. 

For the dominant background, $ZZ\rightarrow\ell^+\ell^-\bar{\nu}\nu$,  determined from simulated samples, 
systematic theoretical 
uncertainties are derived from the generator differences, QCD factorization and renormalization scales, and PDF modeling. 
The largest theoretical uncertainty, the generator difference, is evaluated as the difference in yields calculated 
from samples simulated with SHERPA~1.4.1~\cite{sherpa} and POWHEG~BOX.
The systematic uncertainties associated with the $ZZ$ background are summarized in Table~\ref{tab:zz_unc} 
for each signal region. The luminosity uncertainty is 2.8\% and is derived from beam-separation scans 
performed following the procedure described in Ref.~\cite{lumi2012}.

\begin{table}[htbp]
\begin{center}
\caption{\label{tab:zz_unc} Summary of the systematic uncertainties for the largest background process: 
($ZZ\rightarrow\ell^+\ell^-\bar{\nu}\nu$). Statistical uncertainties are from MC simulation sample size.}
\begin{tabular}{a{2.5cm}x{1cm}x{1cm}x{1cm}x{1cm}}
\hline
\hline
\multicolumn{1}{l}{\multirow{2}{*}{Uncertainty Source}}  & \multicolumn{4}{c}{\MET\ threshold [GeV]} \tn
 & 150 & 250 & 350 & 450 \tn
\hline
Statistical [\%]  & $ 2 $  & $ 6 $  & $ 13 $ & $ 24 $ \tn
Experimental [\%] & $ 3 $  & $ 6 $  & $ 9 $  & $ 8 $ \tn
Theoretical [\%]  & $ 36 $ & $ 37 $ & $ 37 $ & $ 38 $ \tn
Luminosity [\%]   & $ 3 $  & $ 3 $  & $ 3 $  & $ 3 $ \tn
\hline
Total [\%]        & $ 36 $ & $ 38 $ & $ 40 $ & $ 46 $ \tn
\hline
\hline
\end{tabular}
\end{center}
\end{table}

\begin{figure}
\begin{center}
\includegraphics[width=0.45\textwidth]{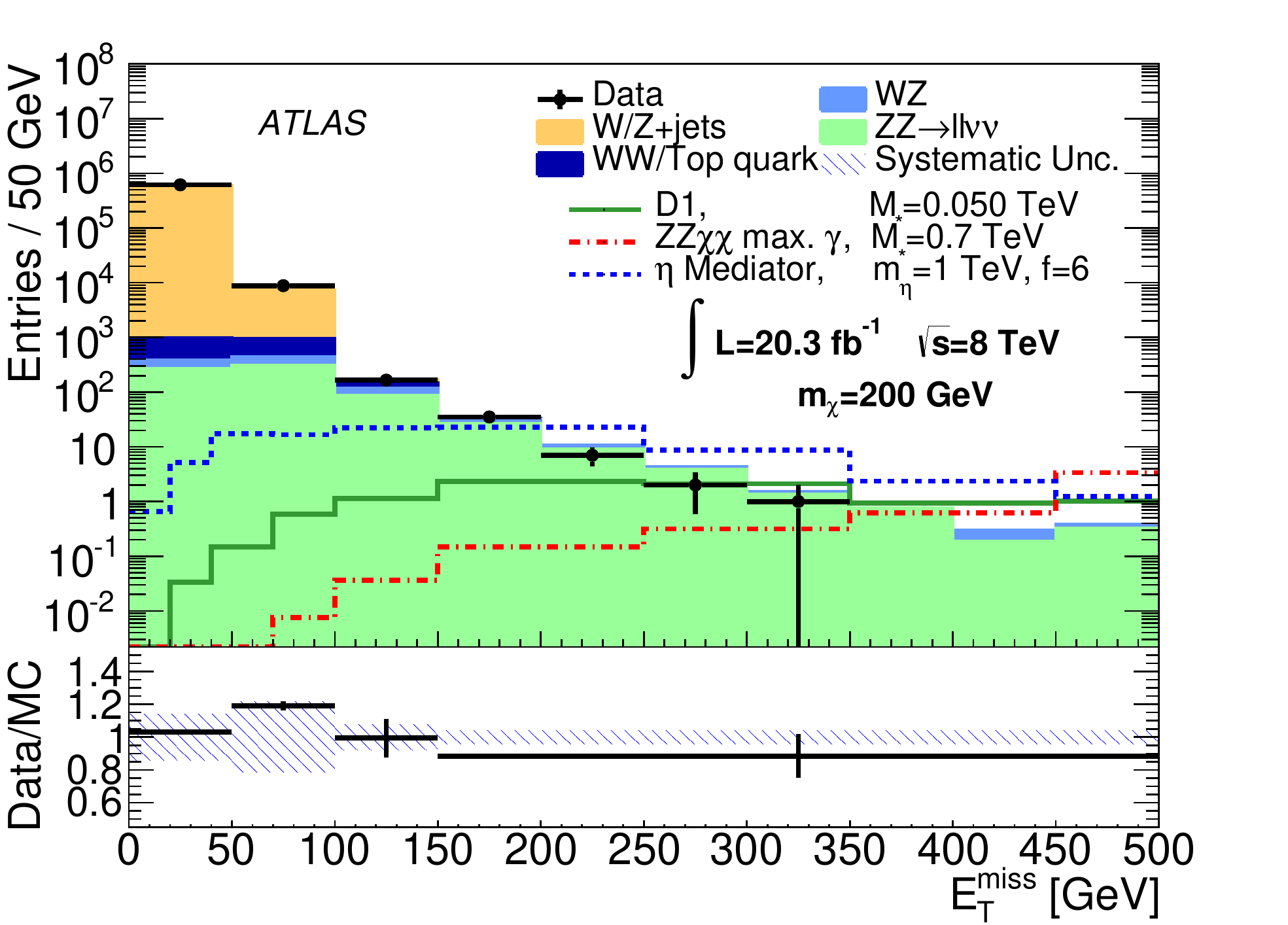}
\caption{\label{fig:met_nm1} 
\MET\ distributions after all event selections other than the \met\
thresholds for the observed data; the expected SM backgrounds taken from simulation;
the hypothetical $pp\rightarrow Z\chi\bar{\chi}$ signals for various values of the mass scale, $M_{\star}$. 
The dark-matter particle mass is $m_{\chi}=200\GeV$. The last bin contains the events with $\MET>450$ GeV. 
The ratio of data to simulated backgrounds is also shown.
The band shows the experimental systematic uncertainties on the ratio.}
\end{center}
\end{figure}

The expected background and observed yields are reported in Table~\ref{tab:exp_yield_limit}. Figure~\ref{fig:met_nm1} 
shows the \MET\ distribution after applying all selection requirements other than the \MET\ thresholds 
for the observed data, the expected SM backgrounds, and the hypothetical $pp\rightarrow Z\chi\bar{\chi}$ signals 
for various values of the mass scale.

\begin{table}[htpb]
\begin{center}
\caption{\label{tab:exp_yield_limit} Observed yields and expected SM backgrounds 
in each signal region. Statistical, systematic, and luminosity uncertainties are added in quadrature 
to give the total background estimate and uncertainties.}
\begin{tabular}{ccccc}
\hline
\hline
\multirow{2}{*}{Process} & \multicolumn{4}{c}{\MET\ threshold [GeV]} \\
 & 150 & 250 & 350 & 450 \\
\hline
$ZZ$ & $41\pm15$ & $6.4\pm2.4$ & $1.3\pm0.5$ & $0.3\pm0.1$ \\
$WZ$ & $8.0\pm3.1$ & $0.8\pm0.4$ & $0.2\pm0.1$ & $0.1\pm0.1$ \\
$WW$, $t\bar{t}$, $Z\rightarrow\tau^+\tau^-$ & $1.9\pm1.4$ & $0^{+0.7}_{-0.0}$ & $0^{+0.7}_{-0.0}$ & $0^{+0.7}_{-0.0}$ \\
$Z$+jets & $0.1\pm0.1$ & -- & -- & -- \\
$W$+jets & $0.5\pm0.3$ & -- & -- & --\\
\hline
Total & $52\pm18$ & $7.2\pm2.8$ & $1.4\pm0.9$ & $0.4^{+0.7}_{-0.4}$ \\
\hline
Data & 45 & 3 & 0 & 0 \\
\hline
\hline
\end{tabular}
\end{center}
\end{table}

No excess over the background is observed. Upper limits on the number of events 
from a new source are calculated employing a frequentist method with a profile
likelihood ratio~\cite{asympt} using the unbinned yields and uncertainties from each \MET\ region. 
The likelihood is a product of a Poisson distribution and Gaussian constraints for the total 
signal and background systematic uncertainties. 
The mean of the Poisson distribution, for either signal and background or background alone, 
includes the effect of varying the nuisance parameters. 

\begin{figure}
\begin{center}
\includegraphics[width=0.45\textwidth]{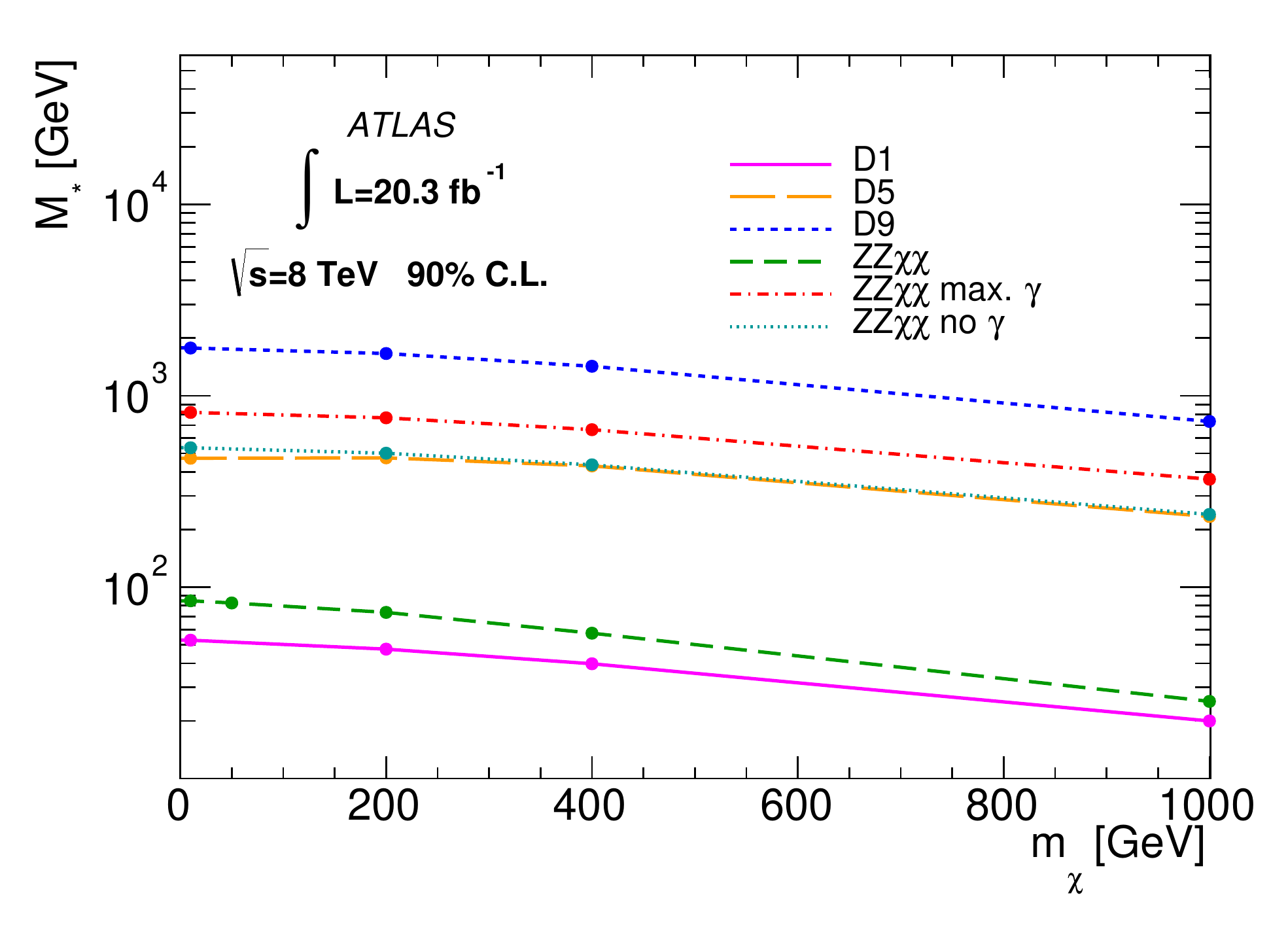}
\end{center}
\caption{\label{fig:collider_limits} Observed 90\%
C.L. lower limits on the mass scale, $M_{\star}$, of considered effective field theories 
as a function of $m_{\chi}$. For 
each operator, the values below the corresponding line are excluded. }
\end{figure}

\begin{figure}
\begin{center}
\includegraphics[width=0.45\textwidth]{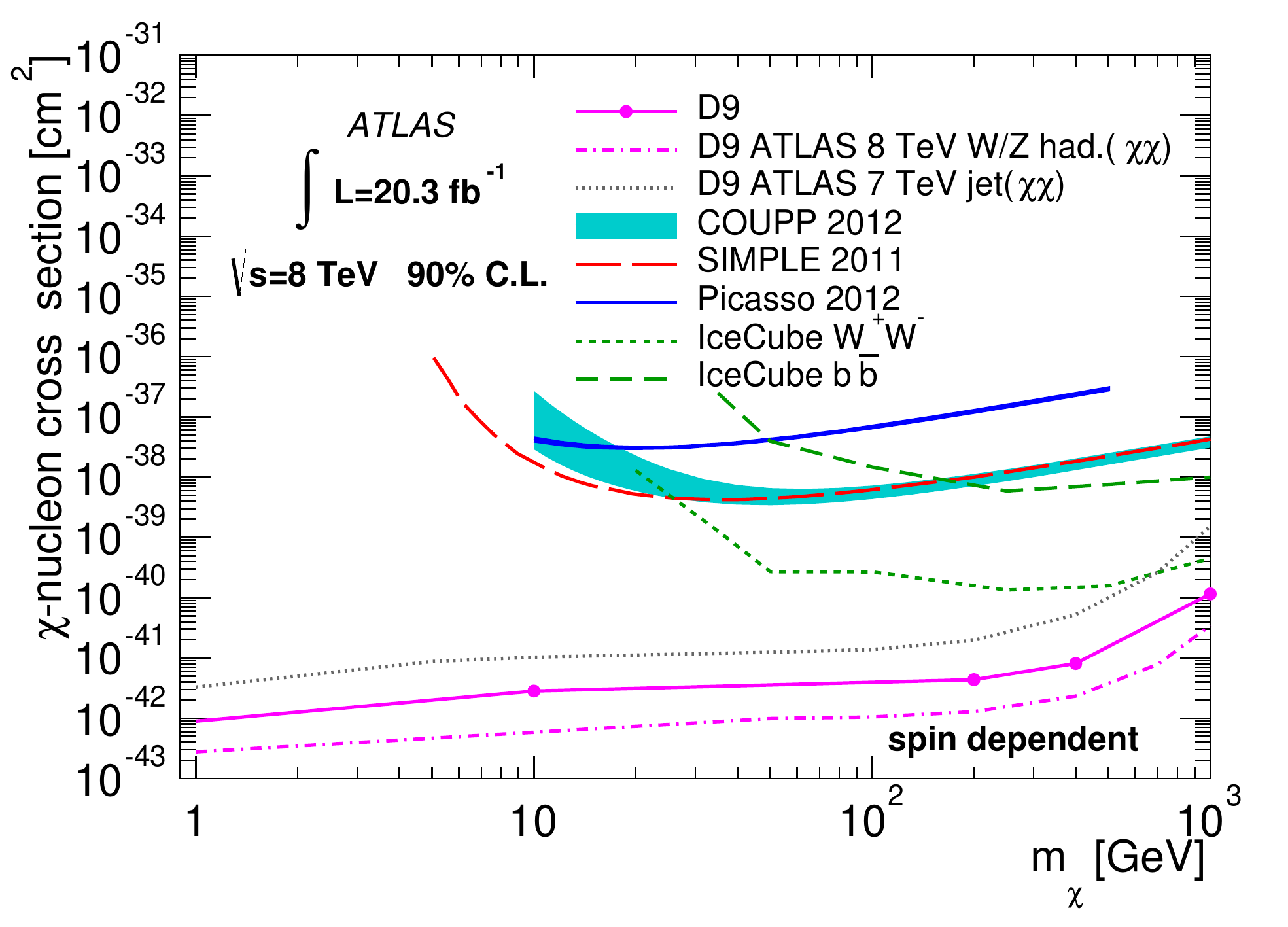}
\end{center}
\caption{\label{fig:dd_sd_limits} Observed 90\% C.L. upper
limits on the $\chi$--nucleon scattering cross section as a function 
of $m_{\chi}$ for the spin-dependent D9 
effective operators mediating the interaction of the dark-matter
particles with the $q\bar{q}$ initial state. The limits 
are compared with results from the published ATLAS hadronically decaying $W/Z$~\cite{atlaswz} and $j$ + $\chi\chi$~\cite{atlasjet} searches, 
COUPP~\cite{coupp2012}, SIMPLE~\cite{simple}, 
PICASSO~\cite{picasso}, and IceCube~\cite{icecube}.
These limits are shown as they are given in the corresponding publications and are only shown for comparison with
the results from this analysis, since they are obtained assuming the interactions are mediated by operators different from those used for the
ATLAS limits.}
\end{figure}

\begin{figure} 
\begin{center} 
\includegraphics[width=0.45\textwidth]{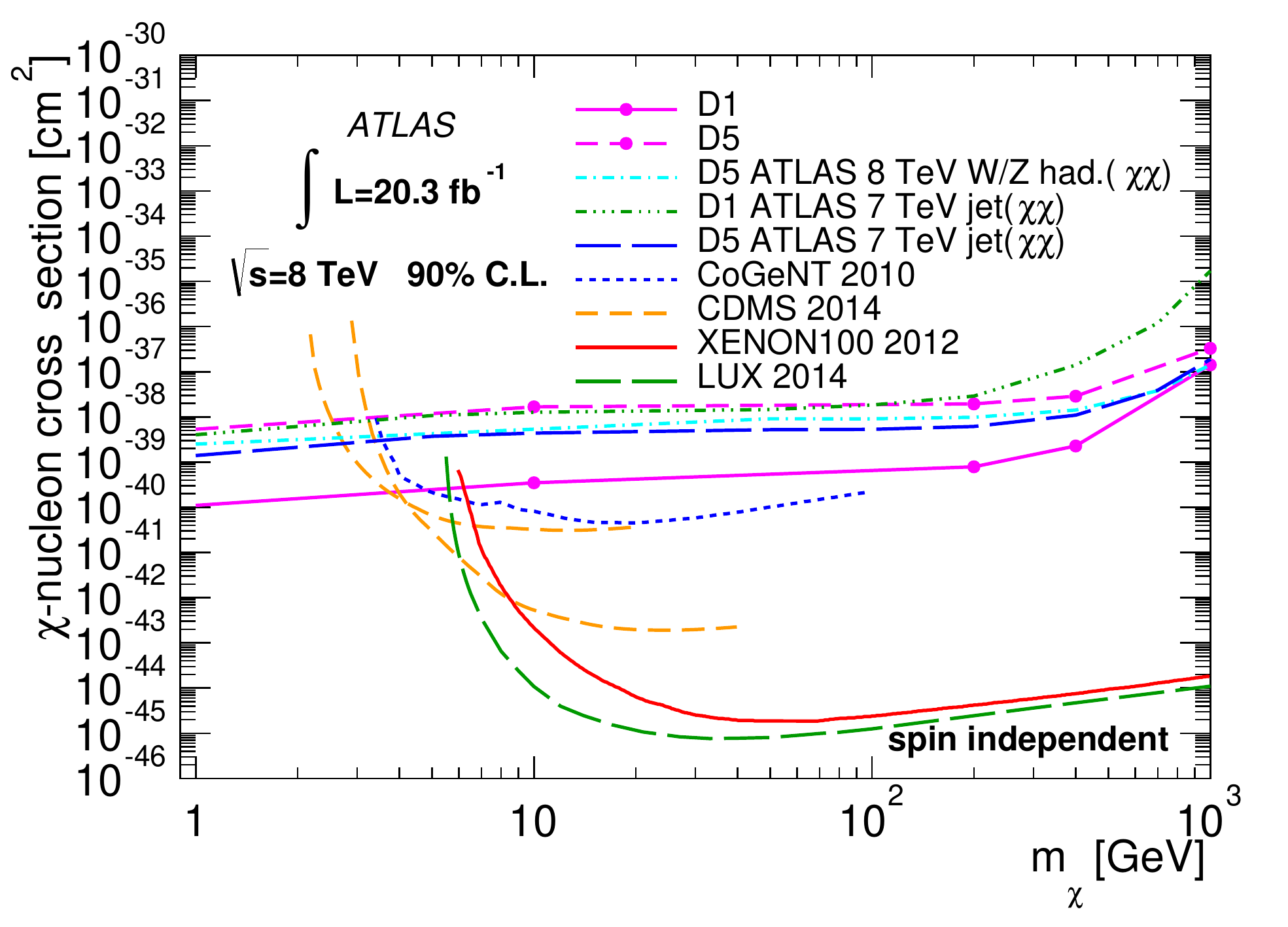} 
\caption{\label{fig:dd_id_limits}Observed 90\% C.L. upper 
limits on the $\chi$--nucleon scattering cross section as a function 
of $m_{\chi}$ for spin-independent
effective operators mediating the interaction of the dark-matter
particles with the $q\bar{q}$ initial state. The limits
are compared with
results from the published ATLAS hadronically decaying $W/Z$~\cite{atlaswz} and $j$ + $\chi\chi$~\cite{atlasjet} searches,
CoGeNT~\cite{cogent}, XENON100~\cite{xenon}, CDMS~\cite{cdmslite,supercdms}, and LUX~\cite{lux}.
These limits are shown as they are given in the corresponding publications and are only shown for comparison with
the results from this analysis, since they are obtained assuming the interactions are mediated by operators different from those used for the
ATLAS limits.
}
\end{center}
\end{figure}

The \MET\ region with the best expected limit is used to calculate the observed limit 
for each operator and mass point. 
Limits on the cross section for $pp\rightarrow Z\chi\bar{\chi}$ production are 
translated into limits on the mass scale of the
effective operators mediating the interaction of the dark-matter particles with the
initial state quarks or the $Z/\gamma^*$ intermediate state. 
This is done using the relation,  
$M_{\star}^{\rm limit}=M_{\star}^{\rm generator} \times \left(\sigma^{\rm generator}/\sigma^{\rm limit}\right)^{1/2p}$, 
where the superscript indicates whether the parameter is a measured limit or calculated using MC simulation, and 
$p$ indicates the power of $(1/M_{\star})$ appearing in the EFT Lagrangian. 
 These limits are shown in Fig.~\ref{fig:collider_limits}.
They are also translated into limits on the $\chi$--nucleon scattering 
cross section using the method in Ref.~\cite{Goodman:2010ku} 
for several effective operators mediating the interaction of the dark-matter
particles with the $q\bar{q}$ initial state, and are compared with other 
experimental results described in Refs.~\cite{coupp2012, simple, picasso, icecube, cogent, xenon, cdmslite, supercdms, lux}. 
These limits, shown at 90\% C.L. in Figs.~\ref{fig:dd_sd_limits} and~\ref{fig:dd_id_limits}, are less stringent than the lower limits for dark-matter candidates 
recoiling against a $W$ or $Z$ boson decaying to hadrons reported in Ref~\cite{atlaswz}. The limits degrade by 13-23\% at 95\% C.L., depending on the \MET\ 
signal region under consideration.

\begin{figure}
\begin{center}
\includegraphics[width=0.45\textwidth]{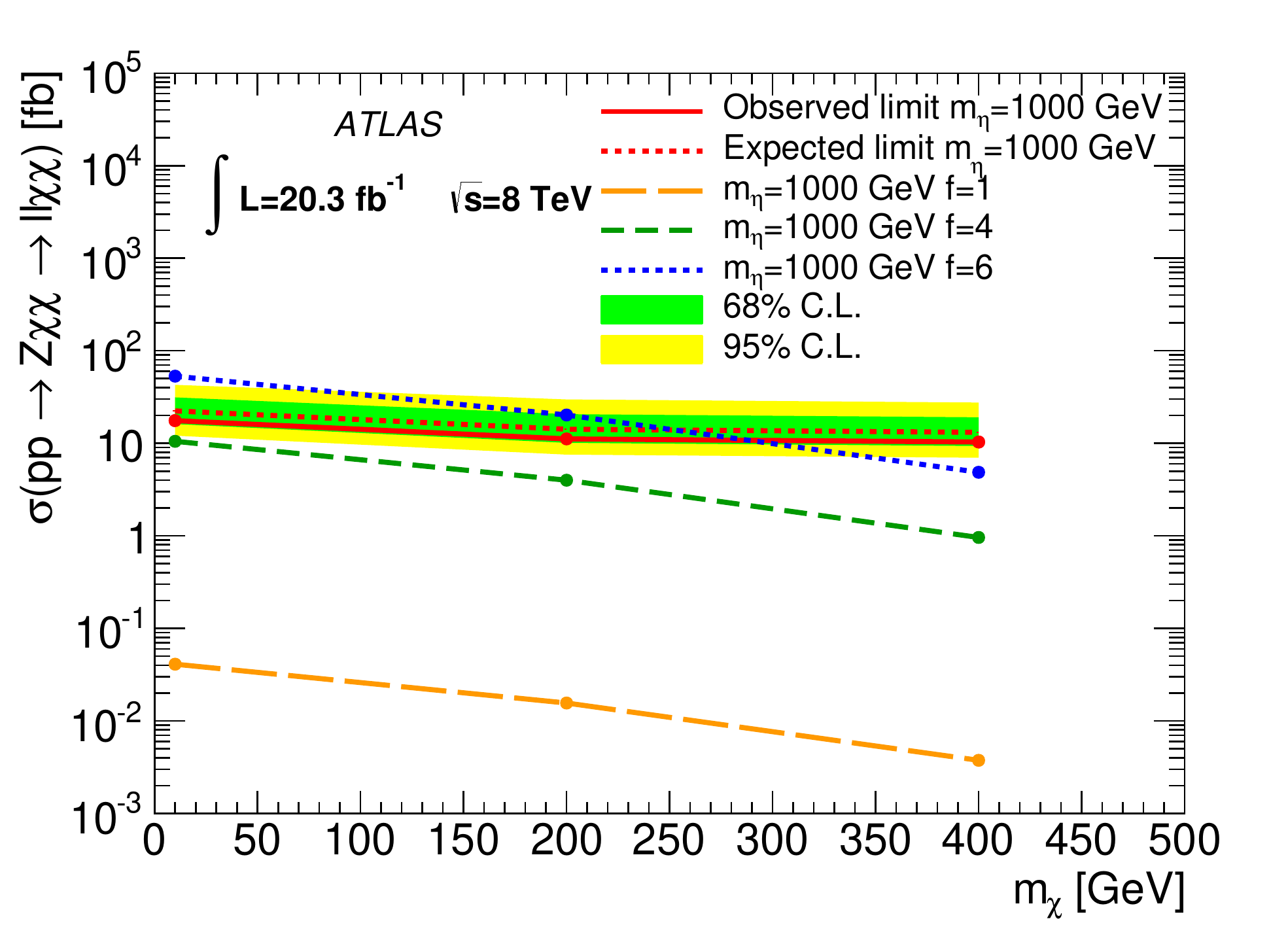}
\end{center}
\caption{\label{fig:mchi_sigma_uv} Observed 95\%
C.L. upper limits on the cross section multiplied by the branching ratio of $Z\rightarrow\ell^+\ell^-$ 
of the scalar-mediator theory as a function of $m_{\chi}$. The observed cross-section limit for a 
scalar-mediator 
mass, $m_{\eta}$, of 1000~GeV is shown. 
Production cross sections predicted from theory are shown for 
$m_{\eta}=1 \TeV$ and for different values of the coupling strength, $f$. } 
\end{figure}

\begin{figure}
\begin{center}
\includegraphics[width=0.45\textwidth]{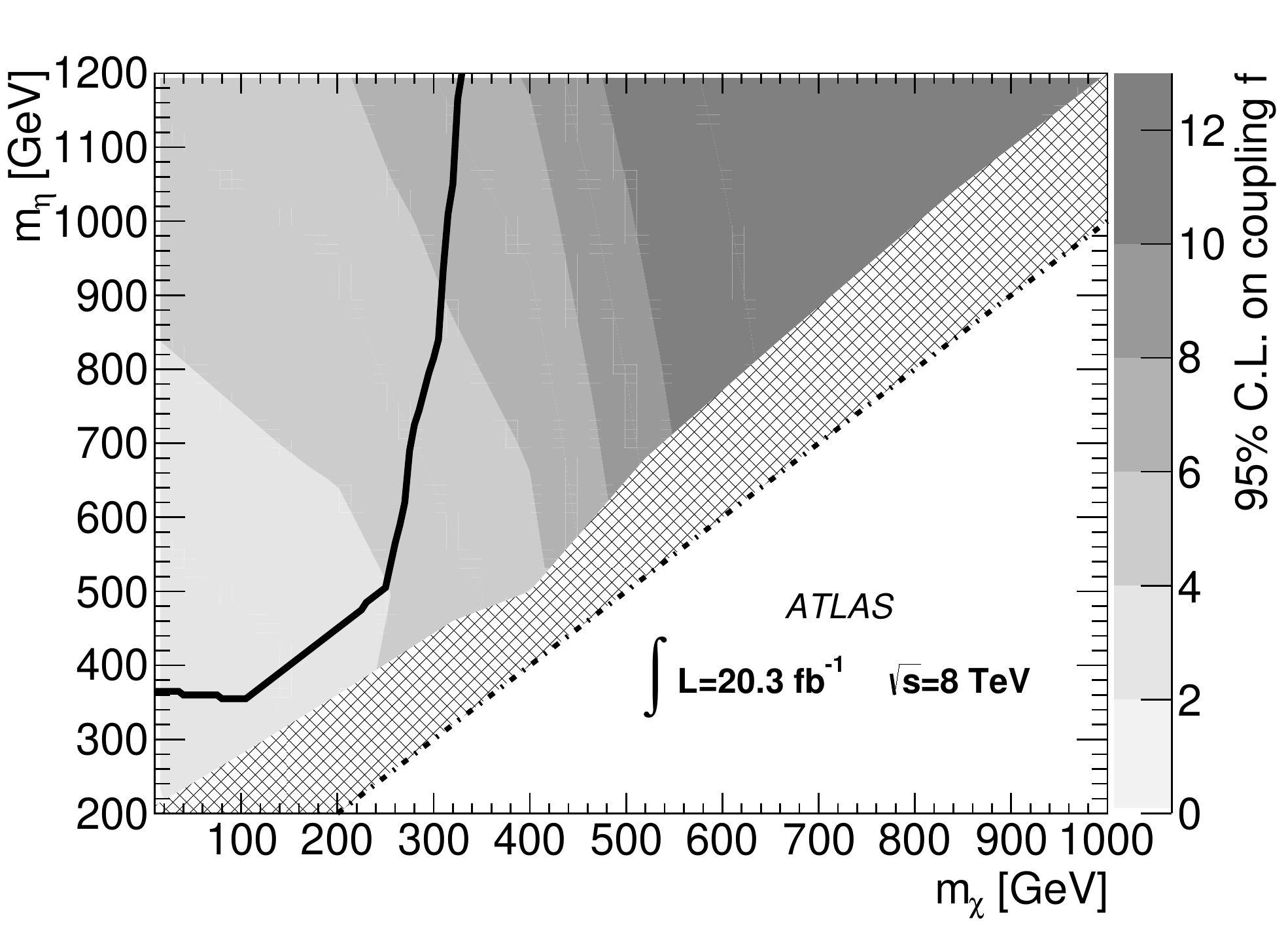}
\end{center}
\caption{\label{fig:f_mchi_vs_meta_uv} Observed 95\%
C.L. upper limits on the coupling constant, $f$, of the scalar-mediator theory  
as a function of $m_{\chi}$ and the mediator mass, $m_{\eta}$. 
The cross-hatching shows the theoretically accessible region outside the range covered by 
this analysis. The white region is phase space beyond the model's validity.
In the excluded region in the upper left-hand corner, 
demarcated by the black line, the lower limit on $f$ from the relic abundance 
calculations based on~\cite{relicAbundance, calculateRelicAbundance}
is greater than the upper limit measured in this analysis. 
}
\end{figure}

A lower limit on the coupling, $f$, of the scalar-mediator model is also calculated based on the WIMP relic  abundance in Ref.~\cite{relicAbundance}
and the expression for the freeze-out temperature from Ref.~\cite{calculateRelicAbundance}. If the relic abundance
lower limit calculated at some mass point $(m_{\chi},m_{\eta})$ is greater than the upper limit measured in this analysis,
that mass point is excluded. Limits on the cross section times branching ratio in the scalar-mediator model are shown in Fig.~\ref{fig:mchi_sigma_uv}, and
limits on $f$ as a function of mediator mass $m_{\eta}$ and $m_{\chi}$, as well as the exclusion region,
are shown in Fig.~\ref{fig:f_mchi_vs_meta_uv}.

Fiducial cross-section limits are calculated in each signal region to complement the limits on specific models. 
The reconstruction efficiency is defined as the ratio of reconstructed events satisfying all the selection criteria 
to the number of generated events within a fiducial region characterized by 
selection requirements at particle level identical to all the requirements on the 
reconstructed dilepton+\MET\ system, where the \met\ is calculated summing over all neutrinos and dark-matter particles.
The acceptance is the ratio of the number of generated events within the fiducial region to the total 
number of generated events. 
In addition, the generated leptons are required to be 
separated by at least $\Delta{R}=0.2$ to match the isolation requirement. The reconstruction efficiency 
ranges from (56.9$\pm$0.9)\% for $ZZ\chi\chi$-max.-$\gamma^{*}$ at $m_{\chi}=1000\GeV$ to 
(77.9$\pm$3.1)\% for D5 at $m_{\chi}=400\GeV$. 
The lowest value of the reconstruction efficiency is used to calculate the fiducial cross-section limits in order 
to be conservative. 
The corresponding acceptances for the previous operators are 
(30.3$\pm$0.5)\% and (2.6$\pm$0.2)\%, respectively, where the uncertainties are purely statistical and the variation 
in the acceptance arises primarily from the different \met\ spectra of the operators. 
The observed and expected upper limits on the fiducial cross section 
are given in Table~\ref{tab:fid_limit}.

\begin{table}[htbp]
\begin{center}
\caption{\label{tab:fid_limit} The observed and expected upper limits 
on the fiducial cross section at 95\% C.L. for each signal region. 
 }
\begin{tabular}{a{3.0cm}x{1cm}x{1cm}x{1cm}x{1cm}}
\hline
\hline
\multicolumn{1}{l}{\multirow{2}{*}{}}  & \multicolumn{4}{c}{\MET\ threshold [GeV]} \tn
 & 150 & 250 & 350 & 450 \tn
\hline
\multicolumn{1}{l}{\multirow{2}{*}{}}  & \multicolumn{4}{c}{Fiducial cross section [fb]} \tn
Expected Limits [fb] & $ 3.0 $  & $ 0.73 $  & $ 0.36 $  & $ 0.27 $ \tn
Observed Limits [fb] & $ 2.7 $  & $ 0.57 $  & $ 0.27 $  & $ 0.26 $ \tn
\hline
\hline
\end{tabular}
\end{center}
\end{table}

In conclusion, a search for the production of dark-matter particles in association with 
a $Z$ boson that decays leptonically in 20.3 fb$^{-1}$ of $pp$ collisions at $\sqrt{s}=$8~\TeV\ is 
presented for three EFT operators where the dark matter interacts directly 
with quarks: D1, D5, and D9. The new limits complement the limits reported in 
other LHC analyses. 
The results are also interpreted using EFT models where the dark matter 
interacts directly with pairs of electroweak bosons. Initial limits are set on 
the mass scale of the $ZZ\chi\chi$ EFT operators describing 
the interaction between dark matter and a $Z$ or $\gamma^{*}$ intermediate state. 
Upper limits are also set on the scattering cross section of dark-matter particles 
with nucleons for effective operators mediating the interaction of 
dark-matter particles with a $q\bar{q}$ initial state, and on a model in which the interaction 
between the dark matter and $Z/\gamma^*$ is mediated by a scalar particle. 

We thank CERN for the very successful operation of the LHC, as well as the
support staff from our institutions without whom ATLAS could not be
operated efficiently.

We acknowledge the support of ANPCyT, Argentina; YerPhI, Armenia; ARC,
Australia; BMWF and FWF, Austria; ANAS, Azerbaijan; SSTC, Belarus; CNPq and FAPESP,
Brazil; NSERC, NRC and CFI, Canada; CERN; CONICYT, Chile; CAS, MOST and NSFC,
China; COLCIENCIAS, Colombia; MSMT CR, MPO CR and VSC CR, Czech Republic;
DNRF, DNSRC and Lundbeck Foundation, Denmark; EPLANET, ERC and NSRF, European Union;
IN2P3-CNRS, CEA-DSM/IRFU, France; GNSF, Georgia; BMBF, DFG, HGF, MPG and AvH
Foundation, Germany; GSRT and NSRF, Greece; ISF, MINERVA, GIF, I-CORE and Benoziyo Center,
Israel; INFN, Italy; MEXT and JSPS, Japan; CNRST, Morocco; FOM and NWO,
Netherlands; BRF and RCN, Norway; MNiSW and NCN, Poland; GRICES and FCT, Portugal; MNE/IFA, Romania; MES of Russia and ROSATOM, Russian Federation; JINR; MSTD,
Serbia; MSSR, Slovakia; ARRS and MIZ\v{S}, Slovenia; DST/NRF, South Africa;
MINECO, Spain; SRC and Wallenberg Foundation, Sweden; SER, SNSF and Cantons of
Bern and Geneva, Switzerland; NSC, Taiwan; TAEK, Turkey; STFC, the Royal
Society and Leverhulme Trust, United Kingdom; DOE and NSF, United States of
America.

The crucial computing support from all WLCG partners is acknowledged
gratefully, in particular from CERN and the ATLAS Tier-1 facilities at
TRIUMF (Canada), NDGF (Denmark, Norway, Sweden), CC-IN2P3 (France),
KIT/GridKA (Germany), INFN-CNAF (Italy), NL-T1 (Netherlands), PIC (Spain),
ASGC (Taiwan), RAL (UK) and BNL (USA) and in the Tier-2 facilities
worldwide.

\bibliography{MonoZPaper}

\begin{thebibliography}{48}%
\makeatletter
\providecommand \@ifxundefined [1]{%
 \@ifx{#1\undefined}
}%
\providecommand \@ifnum [1]{%
 \ifnum #1\expandafter \@firstoftwo
 \else \expandafter \@secondoftwo
 \fi
}%
\providecommand \@ifx [1]{%
 \ifx #1\expandafter \@firstoftwo
 \else \expandafter \@secondoftwo
 \fi
}%
\providecommand \natexlab [1]{#1}%
\providecommand \enquote  [1]{``#1''}%
\providecommand \bibnamefont  [1]{#1}%
\providecommand \bibfnamefont [1]{#1}%
\providecommand \citenamefont [1]{#1}%
\providecommand \href@noop [0]{\@secondoftwo}%
\providecommand \href [0]{\begingroup \@sanitize@url \@href}%
\providecommand \@href[1]{\@@startlink{#1}\@@href}%
\providecommand \@@href[1]{\endgroup#1\@@endlink}%
\providecommand \@sanitize@url [0]{\catcode `\\12\catcode `\$12\catcode
  `\&12\catcode `\#12\catcode `\^12\catcode `\_12\catcode `\%12\relax}%
\providecommand \@@startlink[1]{}%
\providecommand \@@endlink[0]{}%
\providecommand \url  [0]{\begingroup\@sanitize@url \@url }%
\providecommand \@url [1]{\endgroup\@href {#1}{\urlprefix }}%
\providecommand \urlprefix  [0]{URL }%
\providecommand \Eprint [0]{\href }%
\providecommand \doibase [0]{http://dx.doi.org/}%
\providecommand \selectlanguage [0]{\@gobble}%
\providecommand \bibinfo  [0]{\@secondoftwo}%
\providecommand \bibfield  [0]{\@secondoftwo}%
\providecommand \translation [1]{[#1]}%
\providecommand \BibitemOpen [0]{}%
\providecommand \bibitemStop [0]{}%
\providecommand \bibitemNoStop [0]{.\EOS\space}%
\providecommand \EOS [0]{\spacefactor3000\relax}%
\providecommand \BibitemShut  [1]{\csname bibitem#1\endcsname}%
\let\auto@bib@innerbib\@empty
\bibitem [{\citenamefont {{V. Rubin, W. Ford Jr., and N.
  Thonnard}}(1980)}]{rotationcurve}%
  \BibitemOpen
  \bibfield  {author} {\bibinfo {author} {\bibnamefont {{V. Rubin, W. Ford Jr.,
  and N. Thonnard}}},\ }\href {\doibase 10.1086/158003} {\bibfield  {journal}
  {\bibinfo  {journal} {Astrophys. J.}\ }\textbf {\bibinfo {volume} {238}},\
  \bibinfo {pages} {471} (\bibinfo {year} {1980})}\BibitemShut {NoStop}%
\bibitem [{\citenamefont {{E. Komatsu, et al.}}(2011)}]{wmap}%
  \BibitemOpen
  \bibfield  {author} {\bibinfo {author} {\bibnamefont {{E. Komatsu, et
  al.}}},\ }\href {\doibase 10.1088/0067-0049/192/2/18} {\bibfield  {journal}
  {\bibinfo  {journal} {Astrophys. J. Suppl.}\ }\textbf {\bibinfo {volume}
  {192}},\ \bibinfo {pages} {18} (\bibinfo {year} {2011})},\ \Eprint
  {http://arxiv.org/abs/1001.4538} {arXiv:1001.4538 [astro-ph]} \BibitemShut
  {NoStop}%
\bibitem [{\citenamefont {{J. Beringer et al. (Particle Data
  Group)}}(2012)}]{pdg}%
  \BibitemOpen
  \bibfield  {author} {\bibinfo {author} {\bibnamefont {{J. Beringer et al.
  (Particle Data Group)}}},\ }\href {\doibase 10.1103/PhysRevD.86.010001}
  {\bibfield  {journal} {\bibinfo  {journal} {Phys. Rev. D}\ }\textbf {\bibinfo
  {volume} {86}},\ \bibinfo {pages} {010001} (\bibinfo {year}
  {2012})}\BibitemShut {NoStop}%
\bibitem [{\citenamefont {{ATLAS
  Collaboration}}(2013{\natexlab{a}})}]{atlasjet}%
  \BibitemOpen
  \bibfield  {author} {\bibinfo {author} {\bibnamefont {{ATLAS
  Collaboration}}},\ }\href {\doibase 10.1007/JHEP04(2013)075} {\bibfield
  {journal} {\bibinfo  {journal} {J. High Energy Phys.}\ }\textbf {\bibinfo
  {volume} {04}},\ \bibinfo {pages} {075} (\bibinfo {year}
  {2013}{\natexlab{a}})},\ \Eprint {http://arxiv.org/abs/1210.4491}
  {arXiv:1210.4491 [hep-ex]} \BibitemShut {NoStop}%
\bibitem [{\citenamefont {{CMS Collaboration}}(2012{\natexlab{a}})}]{cmsjet}%
  \BibitemOpen
  \bibfield  {author} {\bibinfo {author} {\bibnamefont {{CMS Collaboration}}},\
  }\href {\doibase 10.1007/JHEP09(2012)094} {\bibfield  {journal} {\bibinfo
  {journal} {J. High Energy Phys.}\ }\textbf {\bibinfo {volume} {09}},\
  \bibinfo {pages} {094} (\bibinfo {year} {2012}{\natexlab{a}})},\ \Eprint
  {http://arxiv.org/abs/1206.5663} {arXiv:1206.5663 [hep-ex]} \BibitemShut
  {NoStop}%
\bibitem [{\citenamefont {{ATLAS
  Collaboration}}(2013{\natexlab{b}})}]{atlasphoton}%
  \BibitemOpen
  \bibfield  {author} {\bibinfo {author} {\bibnamefont {{ATLAS
  Collaboration}}},\ }\href {\doibase 10.1103/PhysRevLett.110.011802}
  {\bibfield  {journal} {\bibinfo  {journal} {Phys. Rev. Lett.}\ }\textbf
  {\bibinfo {volume} {110}},\ \bibinfo {pages} {011802} (\bibinfo {year}
  {2013}{\natexlab{b}})},\ \Eprint {http://arxiv.org/abs/1209.4625}
  {arXiv:1209.4625 [hep-ex]} \BibitemShut {NoStop}%
\bibitem [{\citenamefont {{CMS
  Collaboration}}(2012{\natexlab{b}})}]{cmsphoton}%
  \BibitemOpen
  \bibfield  {author} {\bibinfo {author} {\bibnamefont {{CMS Collaboration}}},\
  }\href {\doibase 10.1103/PhysRevLett.108.261803} {\bibfield  {journal}
  {\bibinfo  {journal} {Phys. Rev. Lett.}\ }\textbf {\bibinfo {volume} {108}},\
  \bibinfo {pages} {261803} (\bibinfo {year} {2012}{\natexlab{b}})},\ \Eprint
  {http://arxiv.org/abs/1204.0821} {arXiv:1204.0821 [hep-ex]} \BibitemShut
  {NoStop}%
\bibitem [{\citenamefont {{ATLAS
  Collaboration}}(2014{\natexlab{a}})}]{atlaswz}%
  \BibitemOpen
  \bibfield  {author} {\bibinfo {author} {\bibnamefont {{ATLAS
  Collaboration}}},\ }\href {\doibase 10.1103/PhysRevLett.112.041802}
  {\bibfield  {journal} {\bibinfo  {journal} {Phys. Rev. Lett.}\ }\textbf
  {\bibinfo {volume} {112}},\ \bibinfo {pages} {041802} (\bibinfo {year}
  {2014}{\natexlab{a}})},\ \Eprint {http://arxiv.org/abs/1309.4017}
  {arXiv:1309.4017 [hep-ex]} \BibitemShut {NoStop}%
\bibitem [{\citenamefont {{ATLAS Collaboration}}(2014{\natexlab{b}})}]{ZHinv}%
  \BibitemOpen
  \bibfield  {author} {\bibinfo {author} {\bibnamefont {{ATLAS
  Collaboration}}},\ }\href@noop {} {\bibfield  {journal} {\bibinfo  {journal}
  {Submitted to Phys. Rev. Lett.}\ } (\bibinfo {year} {2014}{\natexlab{b}})},\
  \Eprint {http://arxiv.org/abs/1402.3244} {arXiv:1402.3244 [hep-ex]}
  \BibitemShut {NoStop}%
\bibitem [{\citenamefont {Goodman}\ \emph {et~al.}(2010)\citenamefont
  {Goodman}, \citenamefont {Ibe}, \citenamefont {Rajaraman}, \citenamefont
  {Shepherd}, \citenamefont {Tait},\ and\ \citenamefont {Yu}}]{Goodman:2010ku}%
  \BibitemOpen
  \bibfield  {author} {\bibinfo {author} {\bibfnamefont {J.}~\bibnamefont
  {Goodman}}, \bibinfo {author} {\bibfnamefont {M.}~\bibnamefont {Ibe}},
  \bibinfo {author} {\bibfnamefont {A.}~\bibnamefont {Rajaraman}}, \bibinfo
  {author} {\bibfnamefont {W.}~\bibnamefont {Shepherd}}, \bibinfo {author}
  {\bibfnamefont {T.~M.~P.}\ \bibnamefont {Tait}}, \ and\ \bibinfo {author}
  {\bibfnamefont {H.~B.}\ \bibnamefont {Yu}},\ }\href {\doibase
  10.1103/PhysRevD.82.116010} {\bibfield  {journal} {\bibinfo  {journal} {Phys.
  Rev. D}\ }\textbf {\bibinfo {volume} {82}},\ \bibinfo {pages} {116010}
  (\bibinfo {year} {2010})},\ \Eprint {http://arxiv.org/abs/1008.1783}
  {arXiv:1008.1783 [hep-ph]} \BibitemShut {NoStop}%
\bibitem [{\citenamefont {Beltran}\ \emph {et~al.}(2011)\citenamefont
  {Beltran}, \citenamefont {Hooper}, \citenamefont {Kolb}, \citenamefont
  {Krusberge},\ and\ \citenamefont {Tait}}]{Beltran:2010ww}%
  \BibitemOpen
  \bibfield  {author} {\bibinfo {author} {\bibfnamefont {M.}~\bibnamefont
  {Beltran}}, \bibinfo {author} {\bibfnamefont {D.}~\bibnamefont {Hooper}},
  \bibinfo {author} {\bibfnamefont {E.~W.}\ \bibnamefont {Kolb}}, \bibinfo
  {author} {\bibfnamefont {Z.~A.~C.}\ \bibnamefont {Krusberge}}, \ and\
  \bibinfo {author} {\bibfnamefont {T.~M.~P.}\ \bibnamefont {Tait}},\ }\href
  {\doibase 10.1007/JHEP09(2010)037} {\bibfield  {journal} {\bibinfo  {journal}
  {J. High Energy Phys.}\ }\textbf {\bibinfo {volume} {08}},\ \bibinfo {pages}
  {018} (\bibinfo {year} {2011})},\ \Eprint {http://arxiv.org/abs/1002.4137}
  {arXiv:1002.4137 [hep-ph]} \BibitemShut {NoStop}%
\bibitem [{\citenamefont {Fox}\ \emph {et~al.}(2012)\citenamefont {Fox},
  \citenamefont {Harnik}, \citenamefont {Kopp},\ and\ \citenamefont
  {Tsai}}]{Fox:2011pm}%
  \BibitemOpen
  \bibfield  {author} {\bibinfo {author} {\bibfnamefont {P.~J.}\ \bibnamefont
  {Fox}}, \bibinfo {author} {\bibfnamefont {R.}~\bibnamefont {Harnik}},
  \bibinfo {author} {\bibfnamefont {J.}~\bibnamefont {Kopp}}, \ and\ \bibinfo
  {author} {\bibfnamefont {Y.}~\bibnamefont {Tsai}},\ }\href {\doibase
  10.1103/PhysRevD.85.056011} {\bibfield  {journal} {\bibinfo  {journal} {Phys.
  Rev. D}\ }\textbf {\bibinfo {volume} {85}},\ \bibinfo {pages} {056011}
  (\bibinfo {year} {2012})},\ \Eprint {http://arxiv.org/abs/1109.4398}
  {arXiv:1109.4398 [hep-ph]} \BibitemShut {NoStop}%
\bibitem [{\citenamefont {{L. Carpenter, A. Nelson, C. Shimmin, T. Tait, and D.
  Whiteson}}(2013)}]{monoz7tev}%
  \BibitemOpen
  \bibfield  {author} {\bibinfo {author} {\bibnamefont {{L. Carpenter, A.
  Nelson, C. Shimmin, T. Tait, and D. Whiteson}}},\ }\href {\doibase
  10.1103/PhysRevD.87.074005} {\bibfield  {journal} {\bibinfo  {journal} {Phys.
  Rev. D}\ }\textbf {\bibinfo {volume} {87}},\ \bibinfo {pages} {074005}
  (\bibinfo {year} {2013})},\ \Eprint {http://arxiv.org/abs/1212.3352}
  {arXiv:1212.3352 [hep-ex]} \BibitemShut {NoStop}%
\bibitem [{\citenamefont {Cotta}\ \emph {et~al.}(2013)\citenamefont {Cotta},
  \citenamefont {Hewett}, \citenamefont {Le},\ and\ \citenamefont
  {Rizzo}}]{Cotta:2012nj}%
  \BibitemOpen
  \bibfield  {author} {\bibinfo {author} {\bibfnamefont {R.}~\bibnamefont
  {Cotta}}, \bibinfo {author} {\bibfnamefont {J.~L.}\ \bibnamefont {Hewett}},
  \bibinfo {author} {\bibfnamefont {M.~P.}\ \bibnamefont {Le}}, \ and\ \bibinfo
  {author} {\bibfnamefont {T.~G.}\ \bibnamefont {Rizzo}},\ }\href {\doibase
  10.1103/PhysRevD.88.116009} {\bibfield  {journal} {\bibinfo  {journal} {Phys.
  Rev. D}\ }\textbf {\bibinfo {volume} {88}},\ \bibinfo {pages} {116009}
  (\bibinfo {year} {2013})},\ \Eprint {http://arxiv.org/abs/1210.0525}
  {arXiv:1210.0525 [hep-ph]} \BibitemShut {NoStop}%
\bibitem [{\citenamefont {{H.~M. Lee, M. Park, and V. Sanz}}(2013)}]{axionZ}%
  \BibitemOpen
  \bibfield  {author} {\bibinfo {author} {\bibnamefont {{H.~M. Lee, M. Park,
  and V. Sanz}}},\ }\href {\doibase 10.1007/JHEP03(2013)052} {\bibfield
  {journal} {\bibinfo  {journal} {J. High Energy Phys.}\ }\textbf {\bibinfo
  {volume} {03}},\ \bibinfo {pages} {052} (\bibinfo {year} {2013})},\ \Eprint
  {http://arxiv.org/abs/1212.5647} {arXiv:1212.5647 [hep-ph]} \BibitemShut
  {NoStop}%
\bibitem [{\citenamefont {{G.~Busoni, A.~De Simone, E.~Morgante, and
  A.~Riotto}}(2014)}]{eftValid}%
  \BibitemOpen
  \bibfield  {author} {\bibinfo {author} {\bibnamefont {{G.~Busoni, A.~De
  Simone, E.~Morgante, and A.~Riotto}}},\ }\href {\doibase
  10.1016/j.physletb.2013.11.069} {\bibfield  {journal} {\bibinfo  {journal}
  {Phys. Lett. B}\ }\textbf {\bibinfo {volume} {728}},\ \bibinfo {pages} {412}
  (\bibinfo {year} {2014})},\ \Eprint {http://arxiv.org/abs/1307.2253}
  {arXiv:1307.2253 [hep-ph]} \BibitemShut {NoStop}%
\bibitem [{\citenamefont {{N. Bell, J. Dent, A. Galea, T. Jacques, L. Krauss,
  T. Weiler}}(2012)}]{bell}%
  \BibitemOpen
  \bibfield  {author} {\bibinfo {author} {\bibnamefont {{N. Bell, J. Dent, A.
  Galea, T. Jacques, L. Krauss, T. Weiler}}},\ }\href {\doibase
  10.1103/PhysRevD.86.096011} {\bibfield  {journal} {\bibinfo  {journal} {Phys.
  Rev. D}\ }\textbf {\bibinfo {volume} {86}},\ \bibinfo {pages} {096011}
  (\bibinfo {year} {2012})},\ \Eprint {http://arxiv.org/abs/1209.0231}
  {arXiv:1209.0231 [hep-ph]} \BibitemShut {NoStop}%
\bibitem [{\citenamefont {{ATLAS
  Collaboration}}(2008{\natexlab{a}})}]{atlas_det}%
  \BibitemOpen
  \bibfield  {author} {\bibinfo {author} {\bibnamefont {{ATLAS
  Collaboration}}},\ }\href {\doibase 10.1088/1748-0221/3/08/S08003} {\bibfield
   {journal} {\bibinfo  {journal} {JINST}\ }\textbf {\bibinfo {volume} {3}},\
  \bibinfo {pages} {S08003} (\bibinfo {year} {2008}{\natexlab{a}})}\BibitemShut
  {NoStop}%
\bibitem [{Note1()}]{Note1}%
  \BibitemOpen
  \bibinfo {note} {ATLAS uses a right-handed coordinate system with its origin
  at the nominal interaction point (IP) in the center of the detector and the
  $z$-axis along the beam pipe. The $x$-axis points from the IP to the center
  of the LHC ring, and the $y$-axis points upward. Cylindrical coordinates
  $(r,\phi )$ are used in the transverse plane, $\phi $ being the azimuthal
  angle around the beam pipe. The pseudorapidity is defined in terms of the
  polar angle $\theta $ as $\protect \ensuremath {\etaa }=-\mathop {\mathgroup
  \symoperators ln}\nolimits \mathop {\mathgroup \symoperators tan}\nolimits
  (\theta /2)$.}\BibitemShut {Stop}%
\bibitem [{\citenamefont {{ATLAS
  Collaboration}}(2012{\natexlab{a}})}]{egammanote}%
  \BibitemOpen
  \bibfield  {author} {\bibinfo {author} {\bibnamefont {{ATLAS
  Collaboration}}},\ }\href {\doibase 10.1140/epjc/s10052-012-1909-1}
  {\bibfield  {journal} {\bibinfo  {journal} {Eur. Phys. J. C}\ }\textbf
  {\bibinfo {volume} {72}},\ \bibinfo {pages} {1909} (\bibinfo {year}
  {2012}{\natexlab{a}})},\ \Eprint {http://arxiv.org/abs/1110.3174}
  {arXiv:1110.3174 [hep-ex]} \BibitemShut {NoStop}%
\bibitem [{\citenamefont {{M. Cacciari, G. Salam, and G.
  Soyez}}(2008)}]{ktjet}%
  \BibitemOpen
  \bibfield  {author} {\bibinfo {author} {\bibnamefont {{M. Cacciari, G. Salam,
  and G. Soyez}}},\ }\href {\doibase 10.1088/1126-6708/2008/04/063} {\bibfield
  {journal} {\bibinfo  {journal} {J. High Energy Phys.}\ }\textbf {\bibinfo
  {volume} {04}},\ \bibinfo {pages} {063} (\bibinfo {year} {2008})},\ \Eprint
  {http://arxiv.org/abs/0802.1189} {arXiv:0802.1189 [hep-ph]} \BibitemShut
  {NoStop}%
\bibitem [{\citenamefont {{ATLAS
  Collaboration}}(2008{\natexlab{b}})}]{cluster}%
  \BibitemOpen
  \bibfield  {author} {\bibinfo {author} {\bibnamefont {{ATLAS
  Collaboration}}},\ }\href {https://cds.cern.ch/record/1099735} {\bibfield
  {journal} {\bibinfo  {journal} {ATL-LARG-PUB-2008-002}\ } (\bibinfo {year}
  {2008}{\natexlab{b}})},\ \bibinfo {note}
  {\url{https://cds.cern.ch/record/1099735}}\BibitemShut {NoStop}%
\bibitem [{\citenamefont {{ATLAS
  Collaboration}}(2012{\natexlab{b}})}]{metnote}%
  \BibitemOpen
  \bibfield  {author} {\bibinfo {author} {\bibnamefont {{ATLAS
  Collaboration}}},\ }\href {\doibase 10.1140/epjc/s10052-011-1844-6}
  {\bibfield  {journal} {\bibinfo  {journal} {Eur. Phys. J. C}\ }\textbf
  {\bibinfo {volume} {72}},\ \bibinfo {pages} {1844} (\bibinfo {year}
  {2012}{\natexlab{b}})},\ \Eprint {http://arxiv.org/abs/1108.5602}
  {arXiv:1108.5602 [hep-ex]} \BibitemShut {NoStop}%
\bibitem [{\citenamefont {{T. Melia, P. Nason, R. Rontsch, and G.
  Zanderighi}}(2011)}]{powhegbox}%
  \BibitemOpen
  \bibfield  {author} {\bibinfo {author} {\bibnamefont {{T. Melia, P. Nason, R.
  Rontsch, and G. Zanderighi}}},\ }\href {\doibase 10.1007/JHEP11(2011)078}
  {\bibfield  {journal} {\bibinfo  {journal} {J. High Energy Phys.}\ }\textbf
  {\bibinfo {volume} {11}},\ \bibinfo {pages} {078} (\bibinfo {year} {2011})},\
  \Eprint {http://arxiv.org/abs/1107.5051} {arXiv:1107.5051 [hep-ph]}
  \BibitemShut {NoStop}%
\bibitem [{\citenamefont {{J.~M. Campbell, R.~K. Ellis, and C.
  Williams}}(2011)}]{mcfm}%
  \BibitemOpen
  \bibfield  {author} {\bibinfo {author} {\bibnamefont {{J.~M. Campbell, R.~K.
  Ellis, and C. Williams}}},\ }\href {\doibase 10.1007/JHEP07(2011)018}
  {\bibfield  {journal} {\bibinfo  {journal} {J. High Energy Phys.}\ }\textbf
  {\bibinfo {volume} {07}},\ \bibinfo {pages} {18} (\bibinfo {year} {2011})},\
  \Eprint {http://arxiv.org/abs/1105.0020} {arXiv:1105.0020 [hep-ph]}
  \BibitemShut {NoStop}%
\bibitem [{\citenamefont {{H.-L. Lai, M. Guzzi, J. Huston, Z. Li, and P.~M.
  Nadolsky, et al.}}(2010)}]{ct10pdf}%
  \BibitemOpen
  \bibfield  {author} {\bibinfo {author} {\bibnamefont {{H.-L. Lai, M. Guzzi,
  J. Huston, Z. Li, and P.~M. Nadolsky, et al.}}},\ }\href {\doibase
  10.1103/PhysRevD.82.074024} {\bibfield  {journal} {\bibinfo  {journal} {Phys.
  Rev. D}\ }\textbf {\bibinfo {volume} {82}},\ \bibinfo {pages} {074024}
  (\bibinfo {year} {2010})},\ \Eprint {http://arxiv.org/abs/1007.2241}
  {arXiv:1007.2241 [hep-ph]} \BibitemShut {NoStop}%
\bibitem [{\citenamefont {{T. Sjostrand, S. Mrenna, and P.~Z.
  Skands}}(2008)}]{pythia8}%
  \BibitemOpen
  \bibfield  {author} {\bibinfo {author} {\bibnamefont {{T. Sjostrand, S.
  Mrenna, and P.~Z. Skands}}},\ }\href {\doibase 10.1016/j.cpc.2008.01.036}
  {\bibfield  {journal} {\bibinfo  {journal} {Comput. Phys. Commun.}\ }\textbf
  {\bibinfo {volume} {178}},\ \bibinfo {pages} {852} (\bibinfo {year}
  {2008})},\ \Eprint {http://arxiv.org/abs/0710.3820} {arXiv:0710.3820
  [hep-ph]} \BibitemShut {NoStop}%
\bibitem [{\citenamefont {{ATLAS
  Collaboration}}(2012{\natexlab{c}})}]{pythia_au2}%
  \BibitemOpen
  \bibfield  {author} {\bibinfo {author} {\bibnamefont {{ATLAS
  Collaboration}}},\ }\href {https://cdsweb.cern.ch/record/1474107} {\bibfield
  {journal} {\bibinfo  {journal} {ATL-PHYS-PUB-2012-003}\ } (\bibinfo {year}
  {2012}{\natexlab{c}})},\ \bibinfo {note}
  {\url{https://cdsweb.cern.ch/record/1474107}}\BibitemShut {NoStop}%
\bibitem [{\citenamefont {{P. Golonka and Z. Was}}(2006)}]{photos}%
  \BibitemOpen
  \bibfield  {author} {\bibinfo {author} {\bibnamefont {{P. Golonka and Z.
  Was}}},\ }\href {\doibase 10.1140/epjc/s2005-02396-4} {\bibfield  {journal}
  {\bibinfo  {journal} {Eur. Phys. J. C}\ }\textbf {\bibinfo {volume} {45}},\
  \bibinfo {pages} {97} (\bibinfo {year} {2006})},\ \Eprint
  {http://arxiv.org/abs/0506026} {arXiv:0506026 [hep-ph]} \BibitemShut
  {NoStop}%
\bibitem [{\citenamefont {{ATLAS Collaboration}}(2010)}]{simulation}%
  \BibitemOpen
  \bibfield  {author} {\bibinfo {author} {\bibnamefont {{ATLAS
  Collaboration}}},\ }\href {\doibase 10.1140/epjc/s10052-010-1429-9}
  {\bibfield  {journal} {\bibinfo  {journal} {Eur. Phys. J. C}\ }\textbf
  {\bibinfo {volume} {70}},\ \bibinfo {pages} {823} (\bibinfo {year} {2010})},\
  \Eprint {http://arxiv.org/abs/1005.4568} {arXiv:1005.4568 [hep-ex]}
  \BibitemShut {NoStop}%
\bibitem [{\citenamefont {{S. Agostinelli, et al.}}(2003)}]{geant4}%
  \BibitemOpen
  \bibfield  {author} {\bibinfo {author} {\bibnamefont {{S. Agostinelli, et
  al.}}},\ }\href {\doibase http://dx.doi.org/10.1016/S0168-9002(03)01368-8}
  {\bibfield  {journal} {\bibinfo  {journal} {{Nucl. Instrum. Methods Phys.
  Res., Sect. A}}\ }\textbf {\bibinfo {volume} {506}},\ \bibinfo {pages} {250}
  (\bibinfo {year} {2003})}\BibitemShut {NoStop}%
\bibitem [{\citenamefont {{ATLAS Collaboration}}(2013{\natexlab{c}})}]{ABCD}%
  \BibitemOpen
  \bibfield  {author} {\bibinfo {author} {\bibnamefont {{ATLAS
  Collaboration}}},\ }\href {\doibase 10.1016/j.physletb.2013.01.034}
  {\bibfield  {journal} {\bibinfo  {journal} {Phys. Rev. Lett. B}\ }\textbf
  {\bibinfo {volume} {719}},\ \bibinfo {pages} {299} (\bibinfo {year}
  {2013}{\natexlab{c}})},\ \Eprint {http://arxiv.org/abs/1212.5409}
  {arXiv:1212.5409 [hep-ex]} \BibitemShut {NoStop}%
\bibitem [{\citenamefont {{J. Alwall, M. Herquet, F. Maltoni, O. Mattelaer, and
  T. Stelzer}}(2011)}]{madgraph5}%
  \BibitemOpen
  \bibfield  {author} {\bibinfo {author} {\bibnamefont {{J. Alwall, M. Herquet,
  F. Maltoni, O. Mattelaer, and T. Stelzer}}},\ }\href {\doibase
  10.1007/JHEP06(2011)128} {\bibfield  {journal} {\bibinfo  {journal} {J. High
  Energy Phys.}\ }\textbf {\bibinfo {volume} {06}},\ \bibinfo {pages} {128}
  (\bibinfo {year} {2011})},\ \Eprint {http://arxiv.org/abs/1106.0522}
  {arXiv:1106.0522 [hep-ph]} \BibitemShut {NoStop}%
\bibitem [{\citenamefont {{A.~D. Martin, W.~J. Stirling, R.~S. Thorne, and G.
  Watt}}(2009)}]{mstw2008}%
  \BibitemOpen
  \bibfield  {author} {\bibinfo {author} {\bibnamefont {{A.~D. Martin, W.~J.
  Stirling, R.~S. Thorne, and G. Watt}}},\ }\href {\doibase
  10.1140/epjc/s10052-009-1072-5} {\bibfield  {journal} {\bibinfo  {journal}
  {{Eur. Phys. J. C}}\ }\textbf {\bibinfo {volume} {63}},\ \bibinfo {pages}
  {189} (\bibinfo {year} {2009})},\ \Eprint {http://arxiv.org/abs/0901.0002}
  {arXiv:0901.0002 [hep-ph]} \BibitemShut {NoStop}%
\bibitem [{\citenamefont {{T. Gleisberg, S. Hoeche, F. Krauss, M. Schonherr, S.
  Schumann, et al.}}(2009)}]{sherpa}%
  \BibitemOpen
  \bibfield  {author} {\bibinfo {author} {\bibnamefont {{T. Gleisberg, S.
  Hoeche, F. Krauss, M. Schonherr, S. Schumann, et al.}}},\ }\href {\doibase
  10.1088/1126-6708/2009/02/007} {\bibfield  {journal} {\bibinfo  {journal} {J.
  High Energy Phys.}\ }\textbf {\bibinfo {volume} {02}},\ \bibinfo {pages}
  {007} (\bibinfo {year} {2009})},\ \Eprint {http://arxiv.org/abs/0811.4622}
  {arXiv:0811.4622 [hep-ph]} \BibitemShut {NoStop}%
\bibitem [{\citenamefont {{ATLAS
  Collaboration}}(2013{\natexlab{d}})}]{lumi2012}%
  \BibitemOpen
  \bibfield  {author} {\bibinfo {author} {\bibnamefont {{ATLAS
  Collaboration}}},\ }\href {\doibase 10.1140/epjc/s10052-013-2518-3}
  {\bibfield  {journal} {\bibinfo  {journal} {Eur. Phys. J. C}\ }\textbf
  {\bibinfo {volume} {73}},\ \bibinfo {pages} {2518} (\bibinfo {year}
  {2013}{\natexlab{d}})},\ \Eprint {http://arxiv.org/abs/1302.4393}
  {arXiv:1302.4393 [hep-ex]} \BibitemShut {NoStop}%
\bibitem [{\citenamefont {Cowan}\ \emph {et~al.}(2011)\citenamefont {Cowan},
  \citenamefont {Cranmer}, \citenamefont {Gross},\ and\ \citenamefont
  {Vitells}}]{asympt}%
  \BibitemOpen
  \bibfield  {author} {\bibinfo {author} {\bibfnamefont {G.}~\bibnamefont
  {Cowan}}, \bibinfo {author} {\bibfnamefont {K.}~\bibnamefont {Cranmer}},
  \bibinfo {author} {\bibfnamefont {E.}~\bibnamefont {Gross}}, \ and\ \bibinfo
  {author} {\bibfnamefont {O.}~\bibnamefont {Vitells}},\ }\href {\doibase
  10.1140/epjc/s10052-011-1554-0} {\bibfield  {journal} {\bibinfo  {journal}
  {Eur. Phys. J. C}\ }\textbf {\bibinfo {volume} {71}},\ \bibinfo {pages}
  {1554} (\bibinfo {year} {2011})},\ \Eprint {http://arxiv.org/abs/1007.1727}
  {arXiv:1007.1727 [physics.data-an]} \BibitemShut {NoStop}%
\bibitem [{\citenamefont {{E. Behnke, et al. (COUPP
  Collaboration)}}(2012)}]{coupp2012}%
  \BibitemOpen
  \bibfield  {author} {\bibinfo {author} {\bibnamefont {{E. Behnke, et al.
  (COUPP Collaboration)}}},\ }\href {\doibase 10.1103/PhysRevD.86.052001}
  {\bibfield  {journal} {\bibinfo  {journal} {Phys. Rev. D}\ }\textbf {\bibinfo
  {volume} {86}},\ \bibinfo {pages} {052001} (\bibinfo {year} {2012})},\
  \Eprint {http://arxiv.org/abs/1204.3094} {arXiv:1204.3094 [astro-ph.CO]}
  \BibitemShut {NoStop}%
\bibitem [{\citenamefont {{M. Felizardo, et al. (SIMPLE
  Collaboration)}}(2012)}]{simple}%
  \BibitemOpen
  \bibfield  {author} {\bibinfo {author} {\bibnamefont {{M. Felizardo, et al.
  (SIMPLE Collaboration)}}},\ }\href {\doibase 10.1103/PhysRevLett.108.201302}
  {\bibfield  {journal} {\bibinfo  {journal} {Phys. Rev. Lett.}\ }\textbf
  {\bibinfo {volume} {108}},\ \bibinfo {pages} {201302} (\bibinfo {year}
  {2012})},\ \Eprint {http://arxiv.org/abs/1106.3014} {arXiv:1106.3014
  [astro-ph]} \BibitemShut {NoStop}%
\bibitem [{\citenamefont {{S. Archambault, et al. (PICASSO
  Collaboration)}}(2012)}]{picasso}%
  \BibitemOpen
  \bibfield  {author} {\bibinfo {author} {\bibnamefont {{S. Archambault, et al.
  (PICASSO Collaboration)}}},\ }\href {\doibase 10.1016/j.physletb.2012.03.078}
  {\bibfield  {journal} {\bibinfo  {journal} {Phys. Rev. Lett. B}\ }\textbf
  {\bibinfo {volume} {711}},\ \bibinfo {pages} {153} (\bibinfo {year}
  {2012})},\ \Eprint {http://arxiv.org/abs/1202.1240} {arXiv:1202.1240
  [hep-ex]} \BibitemShut {NoStop}%
\bibitem [{\citenamefont {{M. G. Aartsen, et al. (IceCube
  Collaboration)}}(2013)}]{icecube}%
  \BibitemOpen
  \bibfield  {author} {\bibinfo {author} {\bibnamefont {{M. G. Aartsen, et al.
  (IceCube Collaboration)}}},\ }\href {\doibase 10.1103/PhysRevLett.110.131302}
  {\bibfield  {journal} {\bibinfo  {journal} {Phys. Rev. Lett.}\ }\textbf
  {\bibinfo {volume} {110}},\ \bibinfo {pages} {131302} (\bibinfo {year}
  {2013})},\ \Eprint {http://arxiv.org/abs/1212.4097} {arXiv:1212.4097
  [astro-ph]} \BibitemShut {NoStop}%
\bibitem [{\citenamefont {{C. E. Aalseth, et al. (CoGeNT
  Collaboration)}}(2011)}]{cogent}%
  \BibitemOpen
  \bibfield  {author} {\bibinfo {author} {\bibnamefont {{C. E. Aalseth, et al.
  (CoGeNT Collaboration)}}},\ }\href {\doibase 10.1103/PhysRevLett.106.131301}
  {\bibfield  {journal} {\bibinfo  {journal} {Phys. Rev. Lett.}\ }\textbf
  {\bibinfo {volume} {106}},\ \bibinfo {pages} {131301} (\bibinfo {year}
  {2011})},\ \Eprint {http://arxiv.org/abs/1002.4703} {arXiv:1002.4703
  [astro-ph.CO]} \BibitemShut {NoStop}%
\bibitem [{\citenamefont {{E. Aprile, et al. (XENON100
  Collaboration)}}(2012)}]{xenon}%
  \BibitemOpen
  \bibfield  {author} {\bibinfo {author} {\bibnamefont {{E. Aprile, et al.
  (XENON100 Collaboration)}}},\ }\href {\doibase
  10.1103/PhysRevLett.109.181301} {\bibfield  {journal} {\bibinfo  {journal}
  {Phys. Rev. Lett.}\ }\textbf {\bibinfo {volume} {109}},\ \bibinfo {pages}
  {181301} (\bibinfo {year} {2012})},\ \Eprint {http://arxiv.org/abs/1207.5988}
  {arXiv:1207.5988 [astro-ph.CO]} \BibitemShut {NoStop}%
\bibitem [{\citenamefont {{R.~Agnese, et al.}}(2014{\natexlab{a}})}]{cdmslite}%
  \BibitemOpen
  \bibfield  {author} {\bibinfo {author} {\bibnamefont {{R.~Agnese, et al.}}}
  (\bibinfo {collaboration} {SuperCDMS Collaboration}),\ }\href {\doibase
  10.1103/PhysRevLett.112.041302} {\bibfield  {journal} {\bibinfo  {journal}
  {Phys. Rev. Lett.}\ }\textbf {\bibinfo {volume} {112}},\ \bibinfo {pages}
  {041302} (\bibinfo {year} {2014}{\natexlab{a}})},\ \Eprint
  {http://arxiv.org/abs/1309.3259} {arXiv:1309.3259 [physics.ins-det]}
  \BibitemShut {NoStop}%
\bibitem [{\citenamefont {{R.~Agnese, et
  al.}}(2014{\natexlab{b}})}]{supercdms}%
  \BibitemOpen
  \bibfield  {author} {\bibinfo {author} {\bibnamefont {{R.~Agnese, et al.}}}
  (\bibinfo {collaboration} {SuperCDMS Collaboration}),\ }\href@noop {}
  {\bibfield  {journal} {\bibinfo  {journal} {{Submitted to Phys. Rev. Lett.}}\
  } (\bibinfo {year} {2014}{\natexlab{b}})},\ \Eprint
  {http://arxiv.org/abs/1402.7137} {arXiv:1402.7137 [hep-ex]} \BibitemShut
  {NoStop}%
\bibitem [{\citenamefont {{D.~S.~Akerib, et al.}}(2014)}]{lux}%
  \BibitemOpen
  \bibfield  {author} {\bibinfo {author} {\bibnamefont {{D.~S.~Akerib, et
  al.}}} (\bibinfo {collaboration} {LUX Collaboration}),\ }\href {\doibase
  10.1103/PhysRevLett.112.091303} {\bibfield  {journal} {\bibinfo  {journal}
  {Phys. Rev. Lett.}\ }\textbf {\bibinfo {volume} {112}},\ \bibinfo {pages}
  {091303} (\bibinfo {year} {2014})},\ \Eprint {http://arxiv.org/abs/1310.8214}
  {arXiv:1310.8214 [astro-ph.CO]} \BibitemShut {NoStop}%
\bibitem [{\citenamefont {Steigman}\ \emph {et~al.}(2012)\citenamefont
  {Steigman}, \citenamefont {Dasgupta},\ and\ \citenamefont
  {Beacom}}]{relicAbundance}%
  \BibitemOpen
  \bibfield  {author} {\bibinfo {author} {\bibfnamefont {G.}~\bibnamefont
  {Steigman}}, \bibinfo {author} {\bibfnamefont {B.}~\bibnamefont {Dasgupta}},
  \ and\ \bibinfo {author} {\bibfnamefont {J.~F.}\ \bibnamefont {Beacom}},\
  }\href {\doibase 10.1103/PhysRevD.86.023506} {\bibfield  {journal} {\bibinfo
  {journal} {Phys. Rev. D}\ }\textbf {\bibinfo {volume} {86}},\ \bibinfo
  {pages} {023506} (\bibinfo {year} {2012})},\ \Eprint
  {http://arxiv.org/abs/1204.3622} {arXiv:1204.3622 [hep-ph]} \BibitemShut
  {NoStop}%
\bibitem [{\citenamefont {Bertone}\ \emph {et~al.}(2005)\citenamefont
  {Bertone}, \citenamefont {Hooper},\ and\ \citenamefont
  {Silk}}]{calculateRelicAbundance}%
  \BibitemOpen
  \bibfield  {author} {\bibinfo {author} {\bibfnamefont {G.}~\bibnamefont
  {Bertone}}, \bibinfo {author} {\bibfnamefont {D.}~\bibnamefont {Hooper}}, \
  and\ \bibinfo {author} {\bibfnamefont {J.}~\bibnamefont {Silk}},\ }\href
  {\doibase 10.1016/j.physrep.2004.08.031} {\bibfield  {journal} {\bibinfo
  {journal} {Phys. Rept.}\ }\textbf {\bibinfo {volume} {405}},\ \bibinfo
  {pages} {279} (\bibinfo {year} {2005})},\ \Eprint
  {http://arxiv.org/abs/hep-ph/0404175} {arXiv:hep-ph/0404175 [hep-ph]}
  \BibitemShut {NoStop}%
\end{thebibliography}%

\clearpage
\onecolumngrid
\clearpage
\begin{flushleft}
{\Large The ATLAS Collaboration}

\bigskip

G.~Aad$^{\rm 84}$,
T.~Abajyan$^{\rm 21}$,
B.~Abbott$^{\rm 112}$,
J.~Abdallah$^{\rm 152}$,
S.~Abdel~Khalek$^{\rm 116}$,
O.~Abdinov$^{\rm 11}$,
R.~Aben$^{\rm 106}$,
B.~Abi$^{\rm 113}$,
M.~Abolins$^{\rm 89}$,
O.S.~AbouZeid$^{\rm 159}$
% ...
\bigskip
\\
% ...
$^{11}$ Institute of Physics, Azerbaijan Academy of Sciences, Baku, Azerbaijan\\
% ...
$^{21}$ Physikalisches Institut, University of Bonn, Bonn, Germany\\
% ...
$^{84}$ CPPM, Aix-Marseille Universit{\'e} and CNRS/IN2P3, Marseille, France\\
% ...
$^{89}$ Department of Physics and Astronomy, Michigan State University, East Lansing MI, United States of America\\
% ...
$^{106}$ Nikhef National Institute for Subatomic Physics and University of Amsterdam, Amsterdam, Netherlands\\
% ...
$^{112}$ Homer L. Dodge Department of Physics and Astronomy, University of Oklahoma, Norman OK, United States of America\\
$^{113}$ Department of Physics, Oklahoma State University, Stillwater OK, United States of America\\
% ...
$^{116}$ LAL, Universit{\'e} Paris-Sud and CNRS/IN2P3, Orsay, France\\
% ...
$^{152}$ Institute of Physics, Academia Sinica, Taipei, Taiwan\\
% ...
$^{159}$ Department of Physics, University of Toronto, Toronto ON, Canada\\
% ...
\end{flushleft}

\end{document}